\documentclass[prx,twocolumn,amsmath,amssymb,superscriptaddress,showpacs]{revtex4}

\usepackage[pdftex]{graphics}
\usepackage{graphicx}

\renewcommand{\vec}[1]{{\mathbf #1}}
\newcommand{\gap}{ \Delta}

\begin{document}

\title{Revealing puddles of electrons and holes in compensated topological insulators}

\author{N. Borgwardt}
\affiliation{II. Physikalisches Institut, Universit\"{a}t zu K\"{o}ln, Z\"{u}lpicher Strasse 77, D-50937 K\"{o}ln, Germany}
\author{J. Lux}
\affiliation{Institut f{\"u}r Theoretische Physik, Universit\"{a}t zu K\"{o}ln, Z\"{u}lpicher Strasse 77, D-50937 K\"{o}ln, Germany}
\author{I. Vergara}
\affiliation{II. Physikalisches Institut, Universit\"{a}t zu K\"{o}ln, Z\"{u}lpicher Strasse 77, D-50937 K\"{o}ln, Germany}
\author{Zhiwei Wang}
\affiliation{Institute of Scientific and Industrial Research, Osaka University, Ibaraki, Osaka 567-0047, Japan}
\author{A.A. Taskin}
\affiliation{Institute of Scientific and Industrial Research, Osaka University, Ibaraki, Osaka 567-0047, Japan}
\author{Kouji~Segawa}
\affiliation{Institute of Scientific and Industrial Research, Osaka University, Ibaraki, Osaka 567-0047, Japan}
\author{P.H.M. van Loosdrecht}
\affiliation{II. Physikalisches Institut, Universit\"{a}t zu K\"{o}ln, Z\"{u}lpicher Strasse 77, D-50937 K\"{o}ln, Germany}
\author{Yoichi Ando}
\affiliation{II. Physikalisches Institut, Universit\"{a}t zu K\"{o}ln, Z\"{u}lpicher Strasse 77, D-50937 K\"{o}ln, Germany}
\affiliation{Institute of Scientific and Industrial Research, Osaka University, Ibaraki, Osaka 567-0047, Japan}
\author{A. Rosch}
\affiliation{Institut f{\"u}r Theoretische Physik, Universit\"{a}t zu K\"{o}ln, Z\"{u}lpicher Strasse 77, D-50937 K\"{o}ln, Germany}
\author{M. Gr\"{u}ninger\footnote{grueninger@ph2.uni-koeln.de}}
\affiliation{II. Physikalisches Institut, Universit\"{a}t zu K\"{o}ln, Z\"{u}lpicher Strasse 77, D-50937 K\"{o}ln, Germany}

\date{August 13, 2015}

\begin{abstract}
Three-dimensional topological insulators harbour metallic surface states with exotic properties.
In transport or optics, these properties are typically masked by defect-induced bulk carriers.
Compensation of donors and acceptors reduces the carrier density, but the bulk resistivity remains disappointingly small.
We show that measurements of the optical conductivity in BiSbTeSe$_2$ pinpoint the presence
of electron-hole puddles in the bulk at low temperatures, which is essential for understanding DC bulk transport.
The puddles arise from large fluctuations of the Coulomb potential of donors and acceptors,
even in the case of full compensation.
Surprisingly, the number of carriers appearing within puddles drops rapidly with increasing temperature and almost vanishes around 40\,K.\@
Monte Carlo simulations show that a highly non-linear screening effect arising from thermally activated carriers
destroys the puddles at a temperature scale set by the Coulomb interaction between neighbouring dopants,
explaining the experimental observation semi-quantitatively.
This mechanism remains valid if donors and acceptors do not compensate perfectly.
\end{abstract}

\pacs{72.20.-i, 74.62.Dh, 78.20.-e, 78.30.-j}


\maketitle

\section{Introduction}
\label{sec:intro}

Three-dimensional topological insulators attract significant attention mostly because they feature two-dimensional Dirac fermions
on the surface that possess the peculiar characteristics of spin-momentum locking and topological protection \cite{Hasan10,Qi11,Hasan11,Ando13}.
The existence of such Dirac fermions has been confirmed by surface-sensitive techniques such as angle-resolved photoelectron spectroscopy
or scanning-tunneling microscopy \cite{Xia09,Chen09,Hsieh09,Sanchez14,Alpichshev10,Beidenkopf11}.
However, the exotic phenomena expected in the electromagnetic response of these systems largely remain unexplored to date.
Prominent examples are a  topological magnetoelectric effect related to the quantum Hall effect of the surface states
yielding magnetic monopoles as mirror charges of electric charges \cite{Hasan10,Qi11},
a universal Faraday rotation angle given by the vacuum fine-structure constant $\alpha$ \cite{Qi08,Tse10},
or a universal surface conductance $G(\omega)$\,=\,$\pi e^2/8h$ for energies $\hbar\omega$
larger than twice the Fermi energy $E_F$ \cite{Schmeltzer13,Li13}.
These effects are blurred by a dominant bulk conductivity in real specimen of topological insulators such as
the prototypical binary tetradymites Bi$_2$Te$_3$ and Bi$_2$Se$_3$, which can be categorized as degenerate semiconductors.
Typically, single crystals of these compounds show defect-induced charge carriers with densities
above a few $10^{18}$\,cm$^{-3}$ \cite{Stordeur92,Analytis10,Ren10,LaForge10,Butch10,Eto10,Post13}.

Understanding the defect chemistry allowed for a dramatic reduction of the carrier density \cite{Ando13,Cava13}.
Near-stoichiometric Bi$_2$Se$_3$ exhibits $n$-type conductivity originating from Se vacancies acting as donors,
whereas $p$-type conductivity predominates in Bi$_2$Te$_3$ due to antisite defects.
The most successful route to reduced bulk conductivity aims at two goals in parallel:
reduction of the defect density and compensation of the remaining defects, i.e., $K$\,=\,$N_A/N_D$\,=\,1,
where $N_D$ and $N_A$ denote the densities of donors and acceptors, respectively.
In Bi$_{2-x}$Sb$_x$Te$_{3-y}$Se$_y$, a reduced defect density is achieved by chalcogen order \cite{Taskin11,Ren11}
(see \textit{Methods}, Sec.\ \ref{subsec:samples})
while variation of $x$ and $y$ allows for optimized compensation in combination with the possibility
to tune the energy of the Dirac point with respect to the Fermi energy $E_F$ \cite{Arakane12,Neupane12}.
In BiSbTeSe$_2$, the Dirac point nearly coincides with $E_F$, it thus may serve as a benchmark for the bulk carrier dynamics
at very low carrier concentrations.

For a sample thickness $d \! \lesssim \! 10$\,$\mu$m, the bulk conductance of BiSbTeSe$_2$ is low enough at low temperatures
to be out-weighted by the surface conductance \cite{Taskin11,Xu14,Pan14}.
This allows to observe a hallmark of topological transport, the half-integer quantum Hall effect, at temperatures up to 35\,K \cite{Xu14}.
The bulk resistivity $\rho_b(T)$ never\-theless raises several questions.
At low temperatures, $\rho_b(T)$ of thick samples of Bi$_{2-x}$Sb$_x$Te$_{3-y}$Se$_y$ and also of Bi$_2$Te$_2$Se does not exceed
10\,--\,20\,$\Omega$cm \cite{Ren10,Ren12,Xiong12,Xiong12a,Jia12,Shekhar14,Akrap14,Kushawa14,Ren11,Pan14},
even when the shunting effect of the surface is taken into account \cite{Pan14}.
Above about 100\,K, $\rho_b(T)$ shows activated behavior $\propto \! \exp(E_A/k_B T)$ but the activation energy $E_A$
appears to be substantially smaller than the intrinsic value given by half the gap size, $\gap/2$ \cite{Ren11}.
The bulk conduction mechanism of this important class of materials should be better understood and controlled
for future investigations of novel topological phenomena.

\begin{figure}[tb]
\centering
\includegraphics[width=0.65\columnwidth]{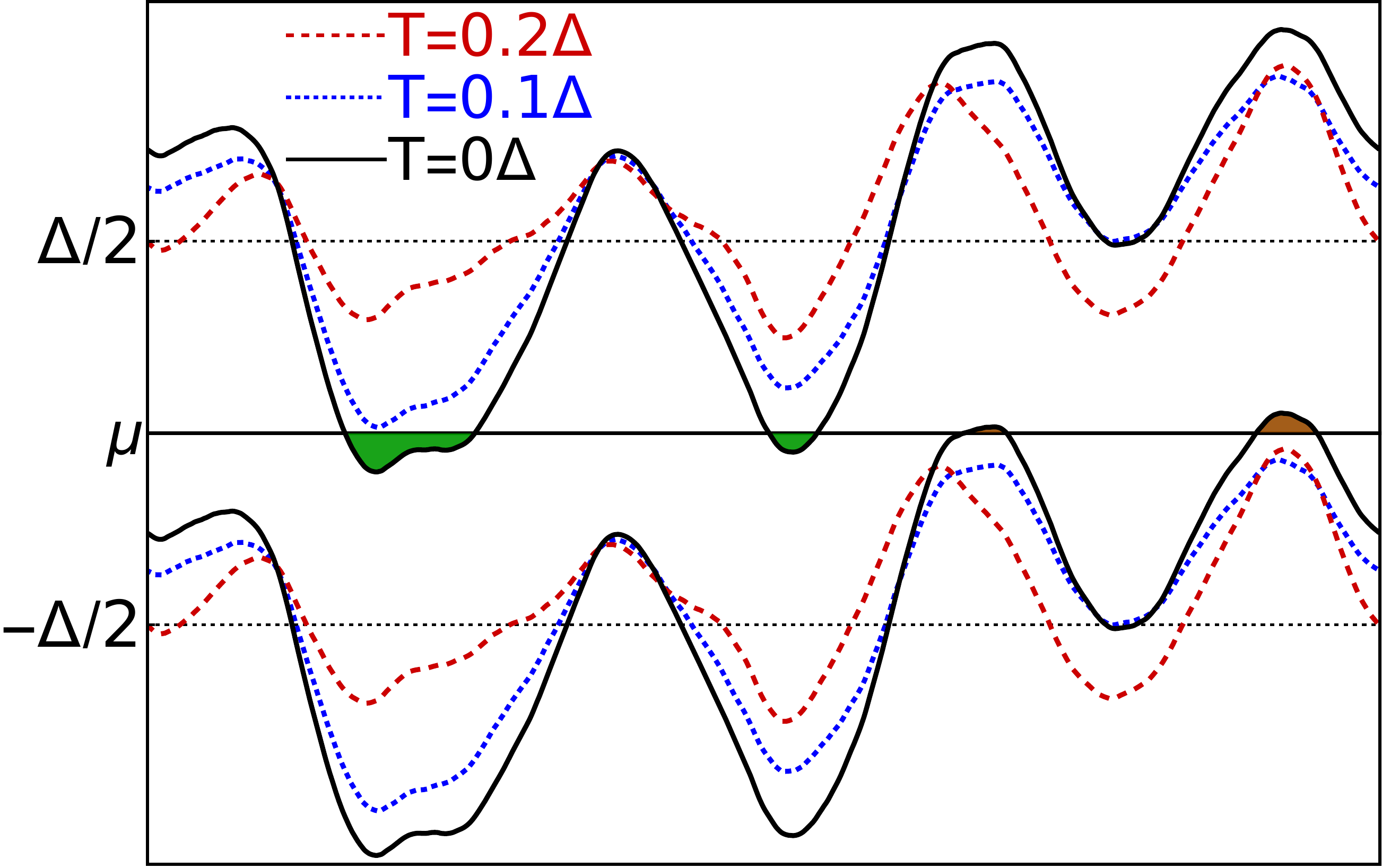}
\includegraphics[width=0.33\columnwidth]{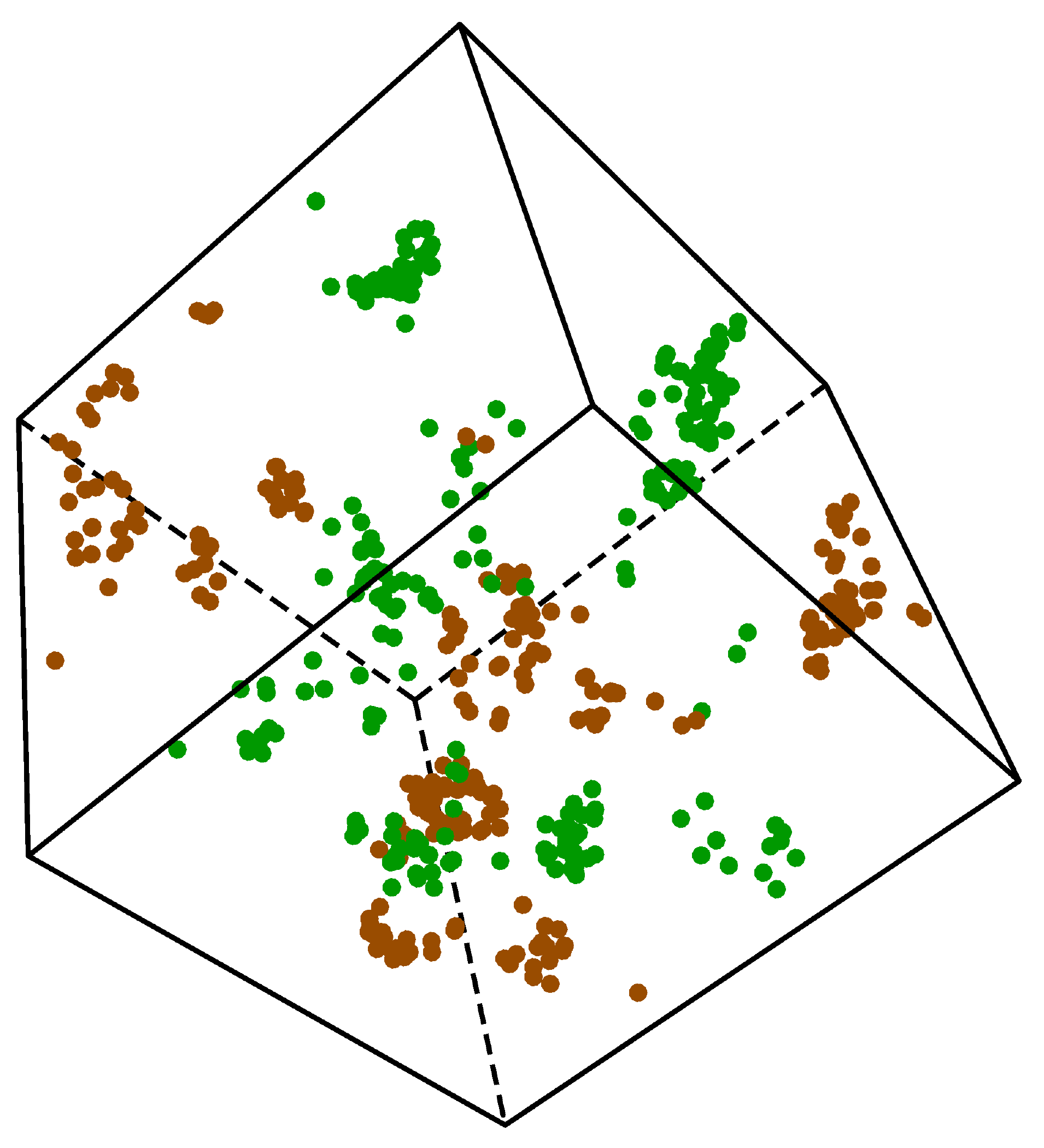}
\caption{Illustration of puddle formation. The left panel depicts the spatial variation of the energies $E_{\pm}(\vec r) = V(\vec r) \pm \gap/2 $
of conduction and valence bands (upper and lower lines) caused by the long-ranged Coulomb potential $V(\vec r)$ arising from
randomly placed donors and acceptors.
At $T$\,=\,0, the bands fluctuate so strongly that the chemical potential $\mu$  is crossed
(shaded areas; data for $\gap/E_c$\,=\,$5$).
This leads to the formation of metallic puddles, i.e., extended regions which are either $n$- or $p$-doped.
An example is shown on the right ($T$\,=\,0, $\gap/E_c$\,=\,10, green/brown: $n/p$-doped).
With increasing temperature, the fluctuations of the potential decrease (dashed lines in left panel) due to screening by
thermally activated carriers thereby suppressing puddle formation.
}
\label{fig:puddles}
\end{figure}

A theoretical explanation of the small activation energy has recently been suggested by Skinner, Chen, and Shklovskii \cite{Skinner12,Skinner13,Chen13}
building on previous work \cite{Shklovskii72}. They considered a perfectly compensated semiconductor ($N_D$\,=\,$N_A$\,=\,$N_{\rm def}$)
with shallow donor and acceptor levels. In such a system, donors give electrons to acceptors, resulting in positively charged donors
and negatively charged acceptors. In this situation, however, the long-range Coulomb interactions necessarily enforce the formation of large puddles, i.e.,
regions in the bulk which are either $p$- or $n$-doped.
The reason is that in a volume of size $R^3$, random fluctuations of the donor and acceptor densities $N_D$ and $N_A$
lead to a typical charge of order $e \sqrt{N_{\rm def} R^3}$ and therefore to a Coulomb potential of order
$e^2 \sqrt{N_{\rm def} R^3}/(4 \pi \varepsilon_0 \varepsilon R)$,
where $\varepsilon$ denotes the dielectric constant and $e$ the elementary charge.
The potential fluctuations grow proportional to $\sqrt{R}$ and become as large as $\Delta/2$ at a length scale
$ R_g=(\Delta /E_c)^2 \, d_{\rm def}/ 8\pi$ \cite{Skinner12}
which is much larger than the average defect distance $d_{\rm def}$\,=\,$N_{\rm def}^{-1/3}$
in the experimentally relevant case $\Delta \! \gg \! E_c$,
where $E_c$\,=\,$e^2/(4 \pi \varepsilon_0 \varepsilon d_{\rm def})$ denotes the Coulomb interaction between neighbouring dopants.
On this length scale $R_g$, the valence and conduction bands are deformed so strongly that they touch and cross the chemical potential,
giving rise to electrically conducting puddles, see Fig.~\ref{fig:puddles}.
In the case of electron puddles, for example, some of the donor states become occupied resulting in neutral donors in this region.
Based on this puddle scenario, Shklovskii and coworkers \cite{Skinner12} find activated behavior of the resistivity,
$\rho_b(T) \propto \! \exp(E_A/k_B T)$, above roughly 40\,K with a \textit{small} activation energy $E_A \approx  0.15\,  \gap$,
consistent with experimental values \cite{Ren11}.
At lower temperatures one expects to observe the famous Efros-Shklovskii law \cite{Efros75} for variable-range hopping,
$\rho_b(T) \propto \! \exp[(T_{\rm ES}/T)^{1/2}]$.
This, however, does not describe the experimentally observed small bulk resistivity at the lowest temperatures.
Nevertheless, the physics of puddle formation is a prime candidate to explain why it is so difficult to reach high bulk resistivities
in compensated topological insulators.

A direct experimental detection of electrically conducting puddles in the bulk of topological insulators is therefore highly desirable.
Surface-sensitive techniques are not ideally suited, as puddles are strongly suppressed close to the metallic surface
which provides an extra screening channel \cite{Skinner13}. Nevertheless, the size of potential fluctuations observed in
scanning tunneling microscopy \cite{Beidenkopf11} appears to be consistent with puddle formation \cite{Skinner13}.

Optical spectroscopy is a bulk-sensitive method ideally suited to detect large conducting regions.
The optical properties of Bi$_2$Te$_3$, Bi$_2$Se$_3$, and of solid solutions thereof were investigated intensively
already half a century ago \cite{Black57,Austin58,Greenaway65,Gobrecht66,Koehler74ph}
in view of their favorable thermoelectric properties \cite{Poudel08,Eibl15}.
Recently, optical data were reported for single crystals of Bi$_2$Te$_2$Se and Bi$_{2-x}$Sb$_x$Te$_{3-y}$Se$_y$
showing reduced carrier density \cite{Akrap12,DiPietro12,Reijnders14,Aleshchenko14,Post15}.
However, the bulk carrier dynamics at very low densities were not addressed in detail.
In particular, these data do not allow to draw conclusions on the presence of puddles.
Here, we give a detailed account of the optical properties of the approximately fully compensated topological insulator BiSbTeSe$_2$
in the infrared range.
We reveal clear signatures of conducting puddles, making use of the recent achievement \cite{Ren11} of very low carrier densities
in BiSbTeSe$_2$ and the sensitivity of transmittance measurements to weak absorption features.
The corresponding spectral weight is strongly temperature dependent at low temperatures.
Based on numerical simulations, we will argue that this temperature dependence is indeed characteristic of
the mechanism of puddle formation by fluctuations of the Coulomb potential.

\section{Experimental Results}
\label{sec:results}

\subsection{Optical spectroscopy}
\label{subsec:optics}

\begin{figure}[tb]
\centering
\includegraphics[width=1.0\columnwidth]{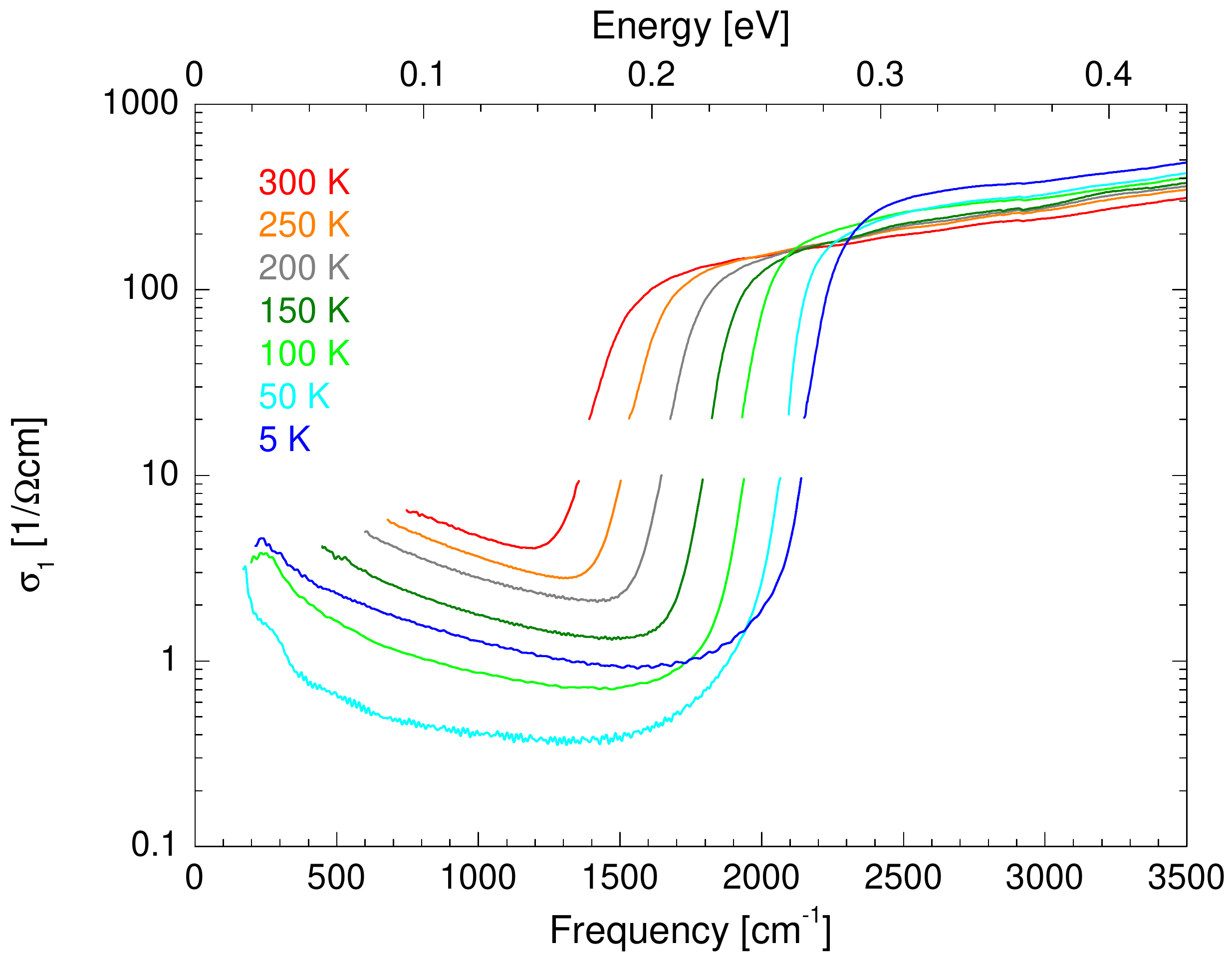}
\caption{Optical conductivity of BiSbTeSe$_2$ on a logarithmic scale.
Weak absorption features below the gap with $\sigma_1(\omega)\! < \! 10$\,$(\Omega$cm$)^{-1}$ were obtained from the transmittance
for a sample thickness of $d$\,=\,102\,$\mu$m, while data with $\sigma_1(\omega) \! > \! 20$\,$(\Omega$cm$)^{-1}$
in the opaque range were derived via a Kramers-Kronig analysis of the reflectivity.
In combination, these data sets give an excellent account of $\sigma_1(\omega)$. }
\label{fig:siglog}
\end{figure}

The complex optical conductivity $\sigma_1(\omega) + i \sigma_2(\omega)$ of single-crystalline BiSbTeSe$_2$
was determined from infrared transmittance and reflectance data which were complemented by ellipsometric measurements
at higher energies (see \textit{Methods}, Sec.\ \ref{subsec:opticalmeas}).
An overview of $\sigma_1(\omega)$ in the infrared range is plotted in Fig.\ \ref{fig:siglog} on a logarithmic scale.
The spectra reveal the steep increase of $\sigma_1(\omega)$ caused by the onset of excitations across the gap $\Delta$.
At 5\,K, we find $\Delta$\,=\,0.26\,eV (2100\,cm$^{-1}$).
At 300\,K, $\Delta$ is reduced by about 40\%, it decreases with a slope of roughly 3.6\,cm$^{-1}$/K.\@
Similar results for the temperature dependence were reported for related topological insulators.
For more details, see \textit{Supplemental Material} \cite{Suppl}.

The main focus of the present study is, however, on the electronic contribution to the optical conductivity below the gap
and its peculiar temperature dependence.
In the temperature window from 40 to 60\,K, $\sigma_1(\omega)$ reaches values as low as 0.3\,$(\Omega$cm$)^{-1}$.
Most remarkably, the temperature dependence of $\sigma_1(\omega)$ is highly non-monotonic.
In the frequency range of about 300\,--\,1100\,cm$^{-1}$, $\sigma_1(\omega)$ is more than three times \textit{larger} at 5\,K than at 50\,K.\@
The rise of $\sigma_1(\omega)$ upon heating above 50\,K agrees with the DC conductivity $\sigma_1(\omega=0)$ measured in transport \cite{Ren11},
but the increase of $\sigma_1(\omega)$ upon cooling below 50\,K strongly deviates from the transport results.
This discrepancy of DC and optical conductivities is the smoking gun for the puddles, as it is a natural consequence of the carrier localization within puddles.

Note that for all temperatures the measured values of  $\sigma_1(\omega)$ below the gap are by far the lowest reported thus far for the entire family
of Bi$_{2-x}$Sb$_x$Te$_{3-y}$Se$_y$.
In Bi$_2$Te$_3$ and Bi$_2$Se$_3$, the Drude contribution of extrinsic carriers with typical densities
$N \! \approx \! 10^{19}$\,cm$^{-3}$ extrapolates to DC values of $\sigma_1(0) \! \approx \! 1000$\,($\Omega$cm)$^{-1}$
\cite{Thomas92,LaForge10,Segura12,DiPietro12,Dordevic13,Post13,Reijnders14,Chapler14}.
In compounds with smaller $N$ such as Bi$_2$Te$_2$Se, impurity absorption bands with peak values of 50\,-\,100\,($\Omega$cm)$^{-1}$
were reported \cite{Akrap12,DiPietro12,Reijnders14}, one to two orders of magnitude larger than the conductivity observed by us.
Such pronounced impurity bands are apparently absent in BiSbTeSe$_2$, in agreement with recent
reflectivity data \cite{Post15}, which were, however, not sensitive enough to reveal
the comparably weak absorption features with $\sigma_1(\omega) \! < \! 10$\,($\Omega$cm)$^{-1}$ observed by us in transmittance.

\subsection{Absence of surface contributions}
\label{subsec:surface}

An important question is whether the spectral weight observed below the gap can be related to the surface states of the topological insulator.
This can, however, be excluded by comparing data for different thicknesses $d$ obtained successively on the same sample (see \textit{Methods}, Sec.\ \ref{sec:exp}).
At each temperature, results for $\sigma_1(\omega)$ for $d$\,=\,102, 130, and 183\,$\mu$m agree very well with each other within the
experimental uncertainty (see Fig.\ S3 in \textit{Supplemental Material} \cite{Suppl}).
This proves the bulk character of the excitations in the investigated frequency range.
Theoretically, one may expect two contributions from the surface state: a Drude peak arising from surface conduction
and \textit{interband} excitations within the Dirac bands.
In BiSbTeSe$_2$, the Fermi level is close to the Dirac point \cite{Arakane12}, giving rise to a small density of surface states.
Moreover, Dirac fermions show a large mobility. The respective narrow Drude peak is located
below the frequency range addressed in our data, in agreement with terahertz data on thin films
of Bi$_2$Se$_3$ and Bi$_{1.5}$Sb$_{0.5}$Te$_{1.8}$Se$_{1.2}$ \cite{Valdes12,Tang13}.
\textit{Interband} excitations within the Dirac bands contribute at higher frequencies.
For $\hbar \omega \! \geq \! 2\,E_F$, a universal conductance $G_0$\,=\,$\pi e^2/8h \approx 1.5\cdot 10^{-5}/\Omega$
has been predicted \cite{Schmeltzer13,Li13}.
For $d$\,=\,100\,$\mu$m, this is equivalent to a bulk conductivity of 0.0015\,($\Omega$cm)$^{-1}$,
which is two orders of magnitude smaller than the lowest values observed in BiSbTeSe$_2$, see Fig.\ \ref{fig:siglog}.
We therefore conclude that all of our observations reflect bulk properties.

\subsection{Electronic contribution to $\sigma_1(\omega)$}
\label{sec:sigma}

\begin{figure}[tb]
\centering
\includegraphics[width=0.95\columnwidth]{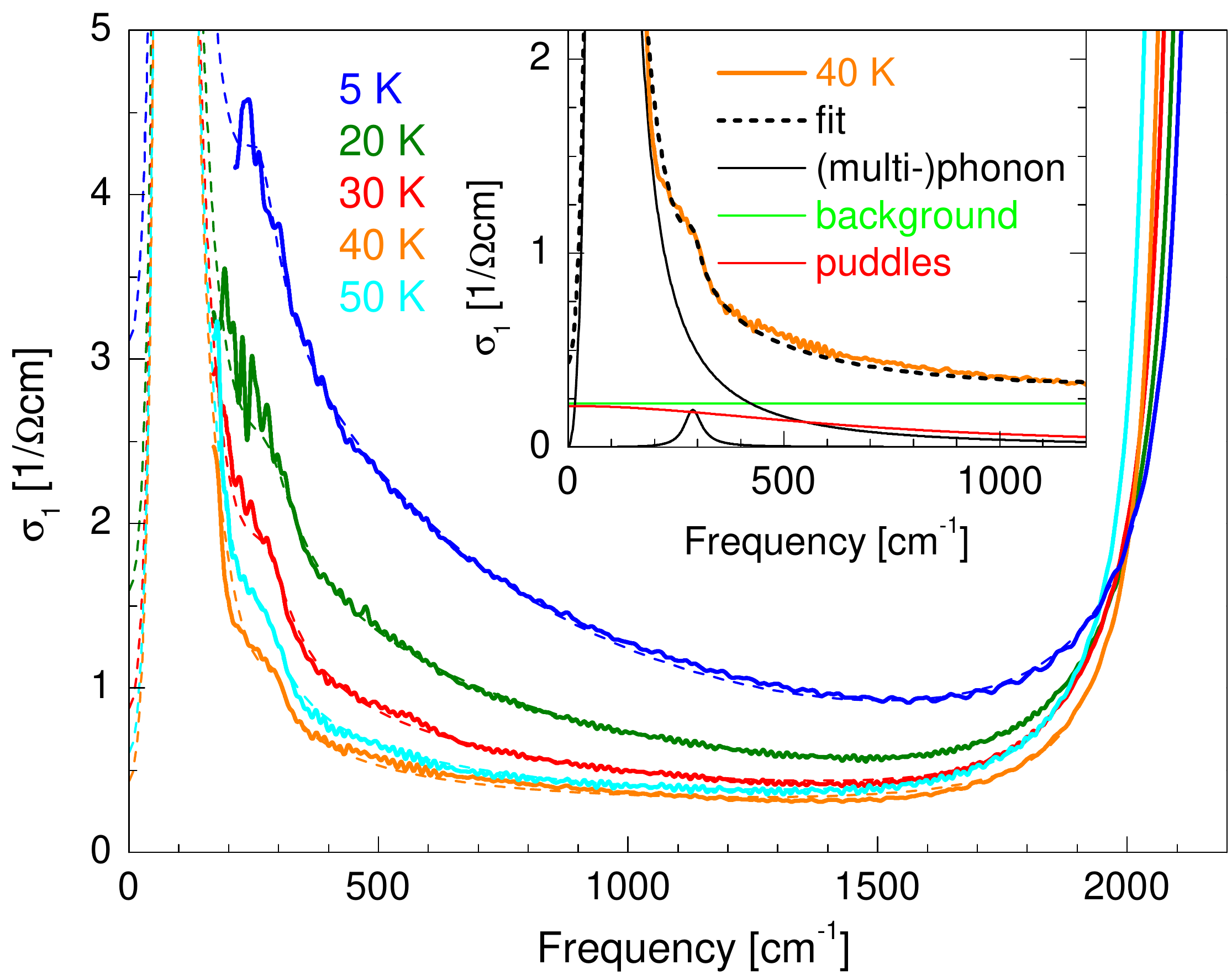}
\includegraphics[width=.95\columnwidth]{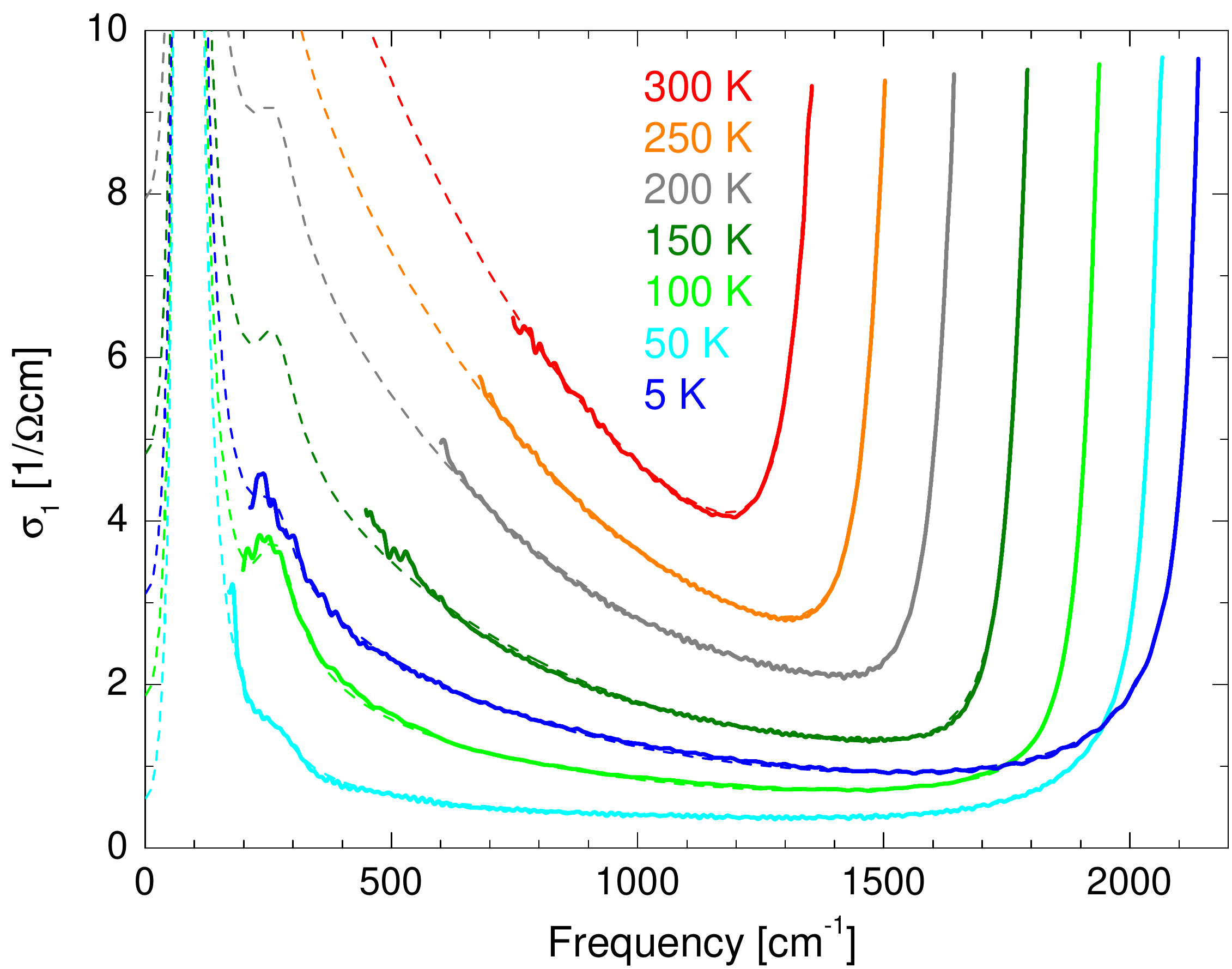}
\caption{Optical conductivity below the gap.
At 40\,--\,50\,K, $\sigma_1(\omega)$ is tiny below the gap. With increasing temperature,
we identify a Drude peak of activated carriers with a strongly temperature-dependent spectral weight (see bottom panel and Fig.~\ref{fig:omegap2})
and a large and approximately temperature-independent scattering rate, $1/\tau \! \approx \! 1.4\cdot 10^{-14}$\,s$^{-1}$.
Most remarkable is the reappearance of low-frequency spectral weight below about 50\,K, which reveals the formation of puddles,
see top panel.
The inset shows a fit of the 40\,K data with four contributions: a phonon at 70\,cm$^{-1}$,
a multi-phonon band at 275\,cm$^{-1}$, a constant background of 0.23\,$(\Omega$cm$)^{-1}$,
and a broad low-frequency band for carriers localized within puddles (red).
The phonon and the background are kept constant in the fits of other temperatures (dashed lines in both panels).
}
\label{fig:sig1}
\end{figure}

Figure \ref{fig:sig1} shows $\sigma_1(\omega)$ on a linear scale for frequencies below the gap.
Several contributions can be identified in this frequency range, see inset of Fig.\ \ref{fig:sig1}.
Below about 150\,cm$^{-1}$, $\sigma_1(\omega)$ is dominated by a phonon contribution with a peak value of the order
of $10^3\,(\Omega$cm$)^{-1}$ \cite{Reijnders14,Post15} which can be identified in the reflectivity data
(see Fig.\ S4 in \textit{Supplemental Material} \cite{Suppl}).
Above 150\,cm$^{-1}$, we find a tiny absorption band extending up to about 350\,cm$^{-1}$ with a peak value of about 1\,($\Omega$cm)$^{-1}$.
Based on the frequency range and the tiny spectral weight, this can be attributed to a multi-phonon contribution,
i.e., two- and three-phonon excitations.
The remaining contributions of electronic origin we fit with a tiny, temperature-independent constant term of about $0.2$\,($\Omega$cm)$^{-1}$ and
a strongly temperature-dependent Drude peak. Well above 50\,K, the interpretation of this feature as a Drude peak of thermally activated carriers
is supported by the absolute value of $\sigma_1(\omega)$, by the peak width, and by the temperature dependence of the spectral weight, as shown below.
The main focus of our study is, however, on the reappearance of spectral weight at low temperatures, which can be attributed
to locally $n$- or $p$-doped puddles. The optical conductivity of such puddles is also expected to be of Drude form for frequencies
above a cut-off $\omega_c$ given by the Thouless energy, determined by the time scale needed to diffuse through a puddle.
Due to the large size of the puddles, the cut-off $\omega_c$ is orders of magnitude smaller than the frequency range investigated by us.
Accordingly, we fit the data using the Drude model also at low temperatures.

In the Drude model, $\sigma_1(\omega)$ depends on the scattering rate $1/\tau$ and the effective carrier density $N_{\rm eff}$\,=\,$N \, m_e/m^*$,
\begin{equation}
\sigma_1(\omega) = \frac{\sigma_1(0)}{1+\omega^2 \tau^2} = \frac{N_{\rm eff}\, e^2 \, \tau /m_e}{1+\omega^2 \tau^2} \, ,
\end{equation}
where $e$ and $m_e$ denote charge and mass of a free electron, respectively, and $m^*$ is the effective band mass.
Well above 50\,K, the fit results for $\sigma_1(0)$ are consistent with DC resistivity data of samples with the same stoichiometry \cite{Ren11}.

\begin{figure}[tb]
\centering
\includegraphics[width=.95\columnwidth]{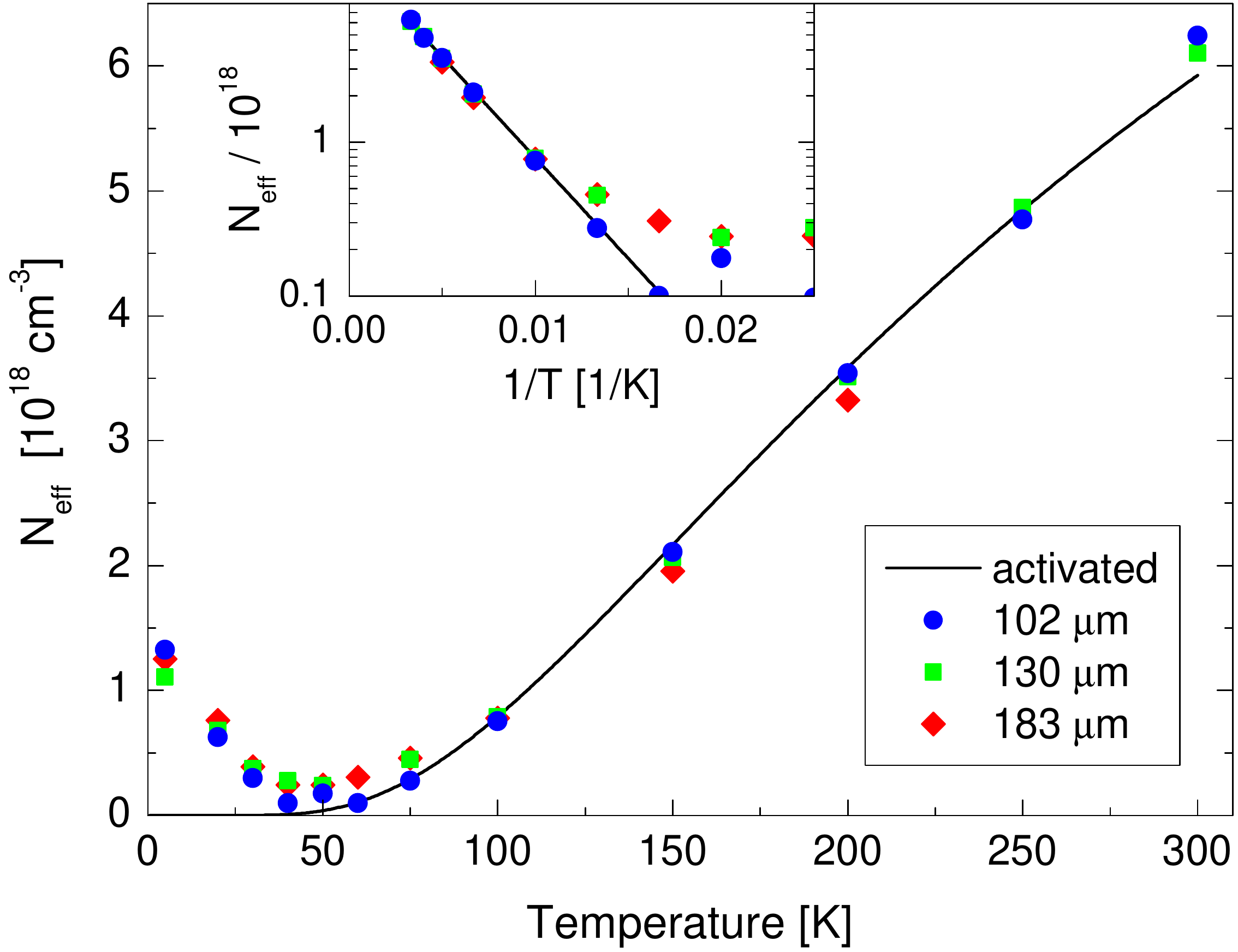}
\caption{Effective carrier density $N_{\rm eff}$.
Symbols depict fitting results for the low-frequency absorption band obtained for different sample thicknesses $d$.
Below about 50\,K, the carriers can be attributed to puddles.
Solid line: activated behavior with an activation energy $E_A$\,=\,26\,meV.\@
Inset: same data on a log scale vs.\ $1/T$.
}
\label{fig:omegap2}
\end{figure}

Comparing our result for $N_{\rm eff}$ at room temperature with Hall-effect data \cite{Ren11}, we find $m^*/m_e$\,=\,0.2, see \textit{Supplemental Material} \cite{Suppl}.
This agrees with results for Bi$_2$Se$_3$, where values between 0.14 to 0.24 were derived from the cyclotron mass of the bulk conduction band
depending on the orientation of the cyclotron orbit \cite{Eto10}.
Using $m^*/m_e$\,=\,0.2, we deduce a carrier density as low as $N \! \approx \! 4\cdot 10^{16}$\,cm$^{-3}$ between 40\,K and 60\,K.\@
The temperature-driven increase of $N_{\rm eff}$ above about 50\,K can be described as activated behavior with an
activation energy $E_A$\,=\,26\,meV $\approx 0.1\,\Delta$,
see inset of Fig.\ \ref{fig:omegap2}. This agrees with $E_A$\,=\,22\,--\,30\,meV derived from transport measurements on BiSbTeSe$_2$
for temperatures above 100\,K \cite{Ren11}.
The small activation energy has been proposed to be a clear signature of strong Coulomb fluctuations \cite{Skinner12,Skinner13}.

From the peak width we obtain $1/\tau \! \approx \! 1.4 \cdot 10^{14}$\,s$^{-1}$ roughly independent of temperature
as expected for a scattering mechanism arising from the random position of defects.
With $m^*/m_e = 0.2$, this corresponds to a mobility $\mu = e\tau/m^*\! \approx \! 70$\,cm$^2$/Vs,
in excellent agreement with the value of 73\,cm$^2$/Vs from Hall data \cite{Ren11} on Bi$_{1.5}$Sb$_{0.5}$Te$_{1.7}$Se$_{1.3}$.
Scattering rates of different compounds are compared in Table \ref{tab:omegap}.
Compensated BiSbTeSe$_2$ shows the smallest carrier density and by far the largest value of $1/\tau$
which supports that defect scattering is dominant.

\begin{table}[t]
\centering
\begin{tabular}{ccccccccc}
  \hline
  compound & \! & $N_{\rm eff}[10^{19}/\text{cm}^{3}$] & \! & $1/\tau[10^{12}$/s] & & $T$ [K] & & Ref.\  \\
  \hline
  BiSbTeSe$_2$   & & 0.02\,/\,0.6  & & 140 & & 50\,/\,300  & & this work \\
  Bi$_2$Te$_2$Se & &   1.9  & & 40 & & 300 & & \cite{Akrap12} \\
  Bi$_2$Se$_3$   & &   2.9; 18  & & 4; 23  & & 6; 300  & & \cite{LaForge10}; \cite{Segura12} \\
  Bi$_2$Te$_3$   & &   33; 46  & & 4.7; 5.6 & & 10; 10 & & \cite{Thomas92};  \cite{Dordevic13} \\
  \hline
\end{tabular}
\caption{Effective carrier densities $N_{\rm eff} = N m_e/m^*$ and scattering rates for different compounds.
Carrier densities from Refs.\ \cite{Akrap12,LaForge10} were calculated from the unscreened plasma frequencies given there.
In Ref.\ \cite{Segura12}, the screened plasma frequency $\omega_p/\sqrt{\varepsilon_\infty}$ is given together
with $\varepsilon_\infty$\,=\,29.5 for Bi$_2$Se$_3$.
}
\label{tab:omegap}
\end{table}

\subsection{Puddles}
\label{sec:puddles}

Our main result is the dramatic reappearance of low-frequency spectral weight at temperatures below 50\,K, see Fig.~\ref{fig:omegap2}.
The charge carriers responsible for this do, however, not contribute to the DC conductivity, and $\sigma_1(\omega)$ at 5\,K
is about an order of magnitude larger than $\sigma_1(\omega=0)$ \cite{Ren11}.
This is consistent with a picture of well separated metallic puddles contributing to $\sigma_1(\omega)$ but not directly to DC transport.

The effective  carrier density amounts to $N_{\rm eff,p} \! \approx \! 1.2\cdot10^{18}$cm$^{-3}$ at 5\,K.\@
Using the value of the effective mass determined at 300\,K, this corresponds to an \textit{average} carrier density
$N_{\rm p} \! \approx \! 2\cdot10^{17}$cm$^{-3}$, which is, however, expected to be distributed in a highly non-uniform way due to puddle formation.
With increasing temperature, the carrier density shows a rapid drop by a factor of 4\,--\,6 at a temperature scale of the order
of 30\,--40\,K, see Fig.~\ref{fig:omegap2}.
Below we will show that this temperature scale has to be identified with the energy scale $E_c$,
which agrees quantitatively with theoretical expectations. Also the average carrier density $N_{\rm p}$ can be explained by our numerical simulations.

Note that  an unconventional -- but much weaker -- temperature dependence of the carrier density has been observed before
in this family of topological insulators.
In Bi$_2$Te$_3$, an unconventional decrease of $N_{\rm eff}$ by up to a factor of 2 has been observed
between 5\,K and 300\,K \cite{Thomas92,Dordevic13,Chapler14}.
For compensated Bi$_2$Te$_2$Se with $N_{\rm eff} \! \approx \! 10^{19}$\,cm$^{-3}$, a non-monotonic behavior
of $N_{\rm eff}$ with a minimum in the range of 50\,K to 150\,K was reported \cite{Aleshchenko14}.
However, $N_{\rm eff}$ in Bi$_2$Te$_2$Se  changes by less than 10\,\% between 5\,K and 50\,K
and by about 20\,\% between 50\,K and 300\,K, whereas we find a drastic change by more than a factor of 10 in BiSbTeSe$_2$,
see Fig.\ \ref{fig:omegap2}. We emphasize that our results are based on samples with very low carrier density in combination
with the enhanced sensitivity for weak absorption features offered by transmittance measurements.

\section{Modelling the formation and destruction of puddles}
\label{sec:model}

Following Skinner, Chen, and Shklovskii \cite{Skinner12}, we use a simple classical electrostatic model to
describe the formation of puddles in a compensated semiconductor.
The model assumes that donors  and acceptors are located at random positions $\vec r_i$ in space.
Their average densities are given by $N_D$ and $N_A$, respectively, with $N_{\rm def}$\,=\,$(N_A+N_D)/2$.
We are mainly interested in the experimentally relevant limit of almost perfect compensation where $K$\,=\,$N_A/N_D$ is close to $1$.
The binding energy of charges to defects is small due to the large dielectric constant
(see Fig.\ S4 in \textit{Supplemental Material} \cite{Suppl}), thus donors and acceptors are shallow with energy levels
very close to $\pm   \gap/2$.
This situation is described by the Hamiltonian
\begin{eqnarray}
H=\sum_i \frac{  \gap}{2} f_i n_i +\frac{1}{2} \sum_{i,j} V_{\vec r_i -\vec r_j} q_i q_j \label{model}
\end{eqnarray}
where $n_i$\,=\,0,1 denotes the number of electrons on a donor ($f_i$\,=\,1) or acceptor ($f_i$\,=\,-1) site.
The charge of a donor (acceptor) amounts to $q_i$\,=\,1 ($q_i$\,=\,-1) if it has donated (accepted) an electron to (from) another defect,
otherwise defects are charge neutral, $q_i$\,=\,0.
The Coulomb potential is supplemented by a short-distance cutoff $a_B$,
$V_{\vec r_i -\vec r_j}$\,=\,$e^2/\{4 \pi \varepsilon_0 \varepsilon (|\vec r_i -\vec r_j|^2+a_B^2)^{1/2}\}$,
which effectively takes into account the finite extent of the wave functions of the shallow impurity states \cite{Skinner12}.
The value of $a_B$ turns out to have little influence \cite{Skinner12} and is set to $a_B$\,=\,$(2/N_{\rm def})^{1/3}$ for all of our simulations.
Expressing all distances in units of the average distance of dopants, $d_{\rm def}=1/N_{\rm def}^{1/3}$,
and all energies in units of the Coulomb interaction between neighbouring dopants, $E_c=e^2/(4 \pi \varepsilon_0 \varepsilon d_{\rm def})$,
all properties of the model depend on $\gap/E_c$, $K$, and $T/E_c$.

The strength of Coulomb interactions and therefore $E_c$ strongly depend on the dielectric constant $\varepsilon$ which is strongly frequency dependent
in BiSbTeSe$_2$ and related compounds.
Below the gap but above the phonons, we find $\varepsilon \! \approx \! 35$, which increases to $\varepsilon \approx 200$ for $\omega \to 0$
due to a huge phonon contribution, see Fig.\ S4 in \textit{Supplemental Material} \cite{Suppl}.
As puddles are static objects around which the highly polarizable ions will adjust their positions, the $\omega \to 0$ value $\varepsilon \approx 200$
should be most relevant for our model and is therefore used in the following.

Besides the donor and acceptor states, no further conduction or valence electron states are taken into account in Eq.\ (\ref{model}).
For a gap of $\Delta/k_B \! \sim \! 3000$\,K, the contribution of intrinsic carriers thermally activated across the gap
can be neglected at low temperatures. Also the intrinsic carrier density within a puddle can be neglected.
This is due to the small effective mass $m^*$ in combination with the small value of the
Fermi energy $E_F \! \sim \! E_c$ within the puddles (see below).
Using $m^*/m_e$\,=\,0.2 and $E_F$\,=\,50\,K for a single spherical band, one obtains an electron density of $10^{17}$\,cm$^{-3}$,
more than an order of magnitude smaller than the typical density of defects.

While the classical model of Eq.~(\ref{model}) is strongly simplified, it is a powerful tool \cite{Shklovskii72,Skinner12,Skinner13,Chen13}
to obtain a semi-quantitative understanding of puddle formation at $T$\,=\,0. We will show below, that it also describes the destruction of puddles with increasing temperature.
Most importantly, the model is sufficiently simple to allow for quantitative numerical simulations both at $T$\,=\,0 and at finite $T$ (see \textit{Methods}, Sec.\ \ref{subsec:simulations}).
We are able to obtain results with only small finite-size effects for values of $  \gap/E_c$ up to 15, see Fig.\ S5 in \textit{Supplemental Material} \cite{Suppl}.
Scaling arguments then allow us to address the experimentally relevant regime of $  \gap/E_c \lesssim 100$ (see Sec.\ \ref{sec:disc}).

\begin{figure}[tb]
\begin{center}
 \includegraphics[width=0.6 \columnwidth]{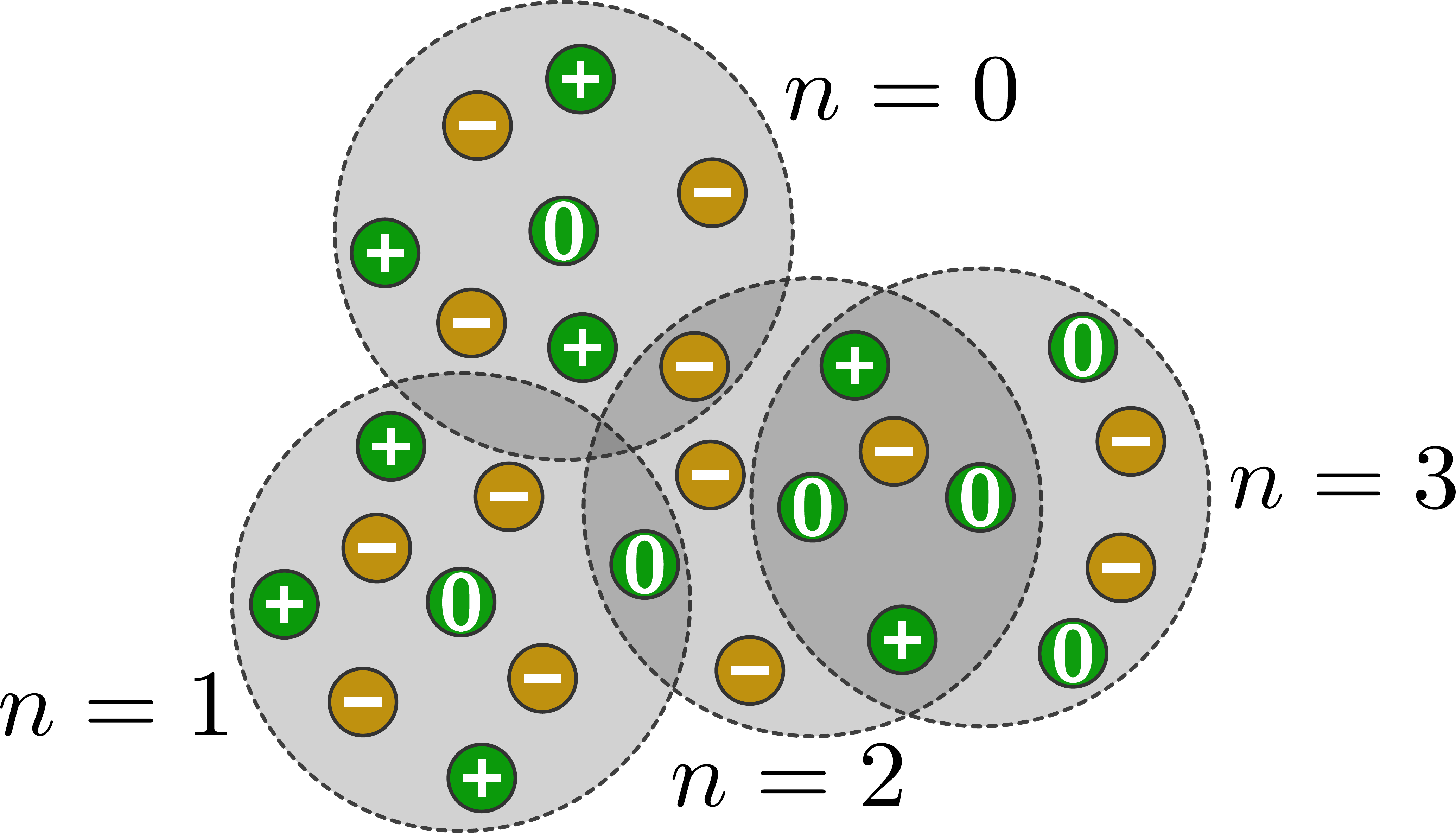}
\includegraphics[width=1.\columnwidth]{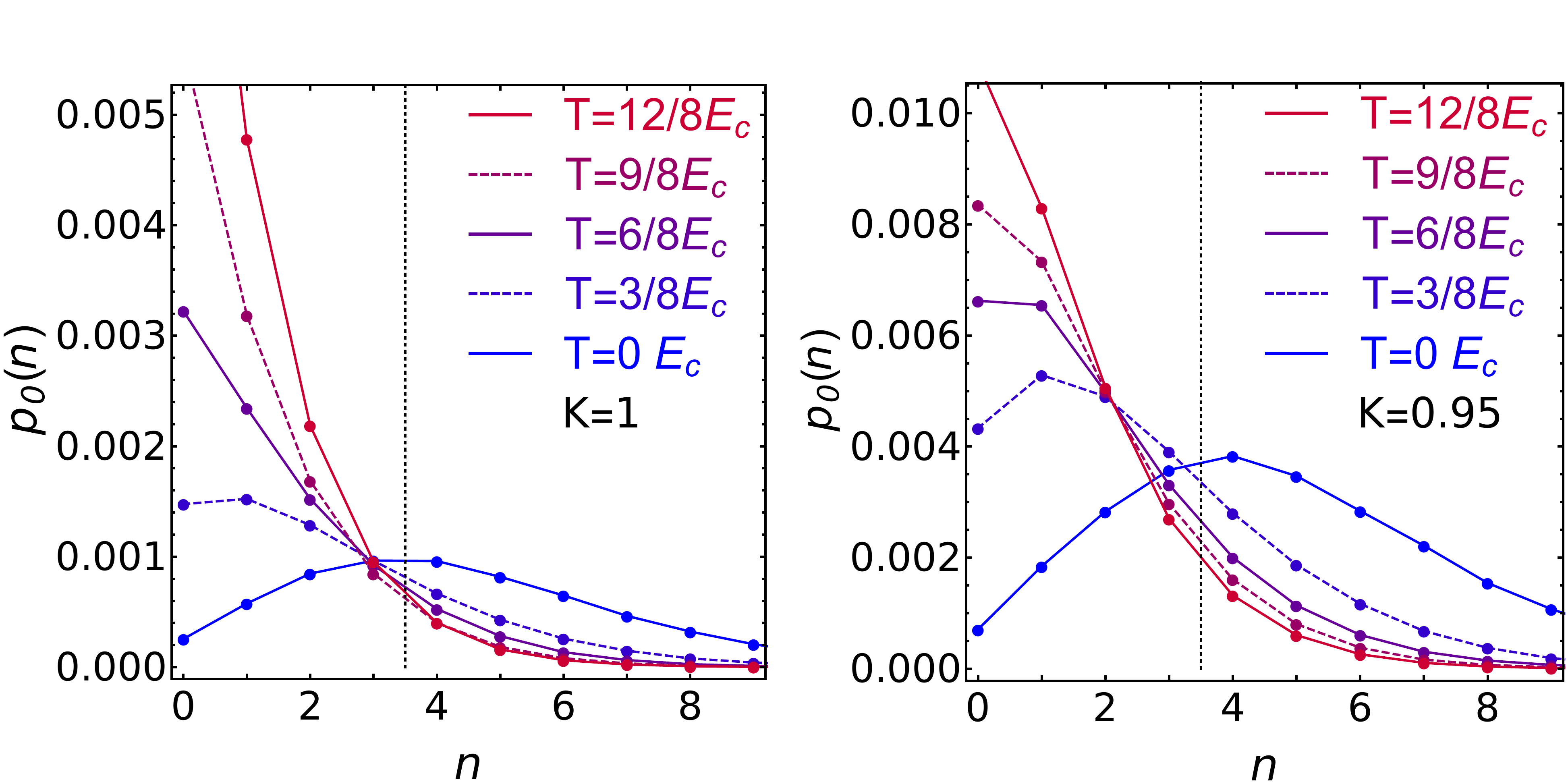}
\end{center}
\caption{Identification of puddles.
Top panel: Schematic picture of an electron puddle in a region where the density of donors (green circles) is larger than the density of acceptors (yellow circles).
The symbols $+,-$, and 0 indicate the dopant charge. To identify a puddle, we consider a sphere of radius  $r_0=1.42 \, d_{\rm def}$ around each neutral dopant.
Then, we count the number $n$ of neutral dopants of the same kind within each sphere to obtain the distribution function $p_0(n)$
(normalized by the total number of dopants).
This is shown in the lower panel for $\gap/E_c$\,=\,15 (left: perfect compensation, $K$\,=\,1; right: $K$\,=\,0.95) and various temperatures.
We identify the fraction $p_p$ of dopants located well within a puddle with $p_p=\sum_{n\ge 4} p_0(n)$, i.e.,
considering all neutral dopants with at least four neutral neighbours (cf.\ vertical dashed lines).
At $T$\,=\,0 (solid blue line), most neutral dopants are organized in puddles,
having a substantial number of neutral neighbours.  With increasing $T$, the number of neutral dopants with many
neutral neighbours drops for $T\lesssim E_c$, which is a clear sign for the destruction of puddles (see Fig.\ \ref{fig:scaling2}).
The number of isolated neutral dopants with no neutral neighbours rises instead.
}
\label{fig:puddle}
\end{figure}

\begin{figure}[tb]
\centering
\includegraphics[width=0.75 \columnwidth]{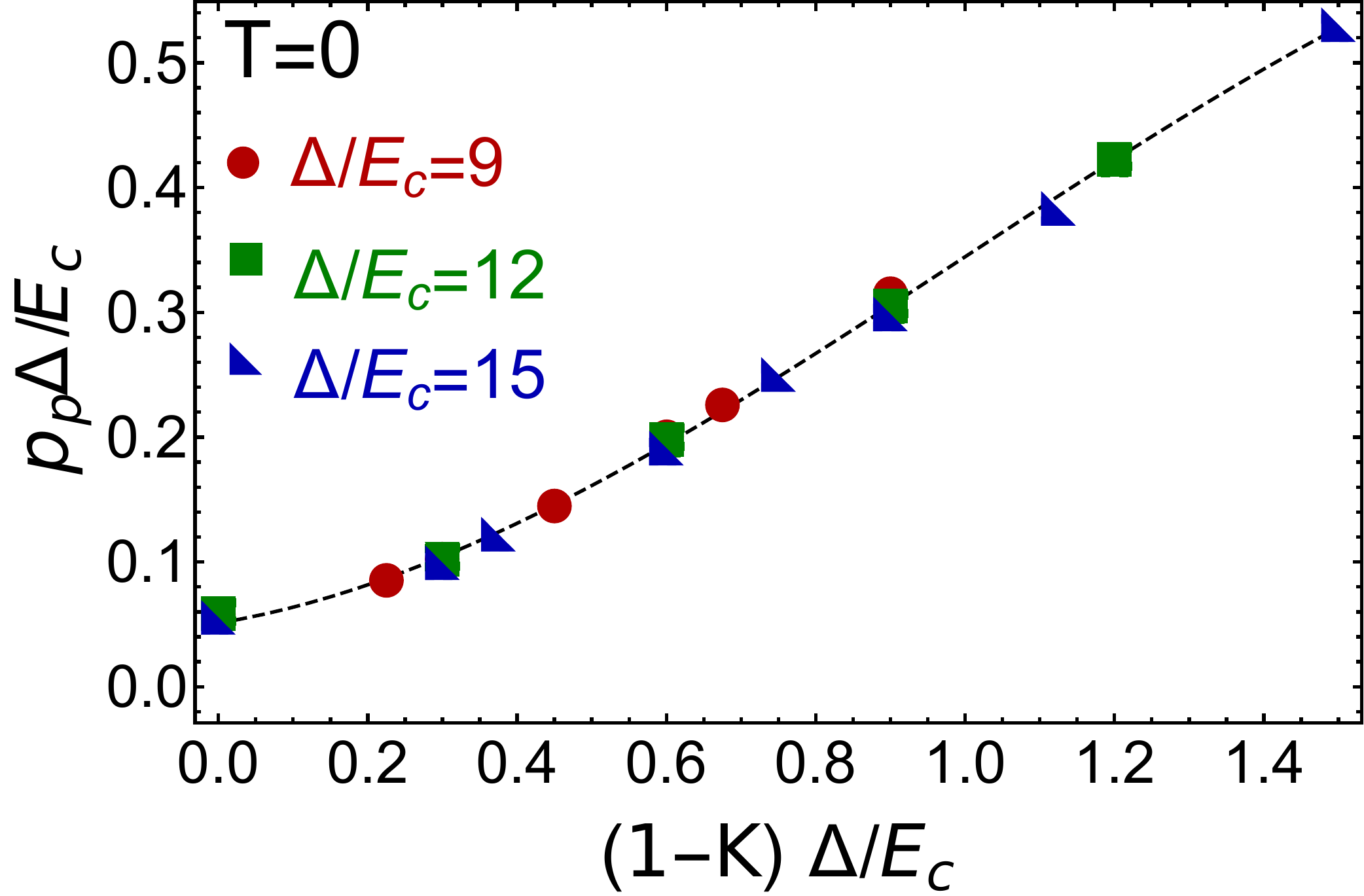}
\caption{Puddle formation and doping.
At $T=0$, the fraction $p_p=\sum_{n\ge 4} p_0(n)$ (see Fig.~\ref{fig:puddle}) of dopants organized in puddles rises strongly
in systems with small deviations $|1-K|\ll 1$ from perfect compensation.
The scaling plot of $p_p \, \Delta/E_c$ as function of $(1-K) \, \Delta/E_c =\frac{N_D-N_A}{N_D}\frac{\Delta}{E_c}$
for different values of $\gap/E_c$ shows that $p_p \approx 0.06\,E_c/\gap$
for perfect compensation, while $p_p \approx 0.33\, (1-K)$ for $1-K \gtrsim 0.3\, E_c/\gap$.
}
\label{fig:scaling1}
\end{figure}

\begin{figure}[tb]
\centering
\includegraphics[width=\columnwidth]{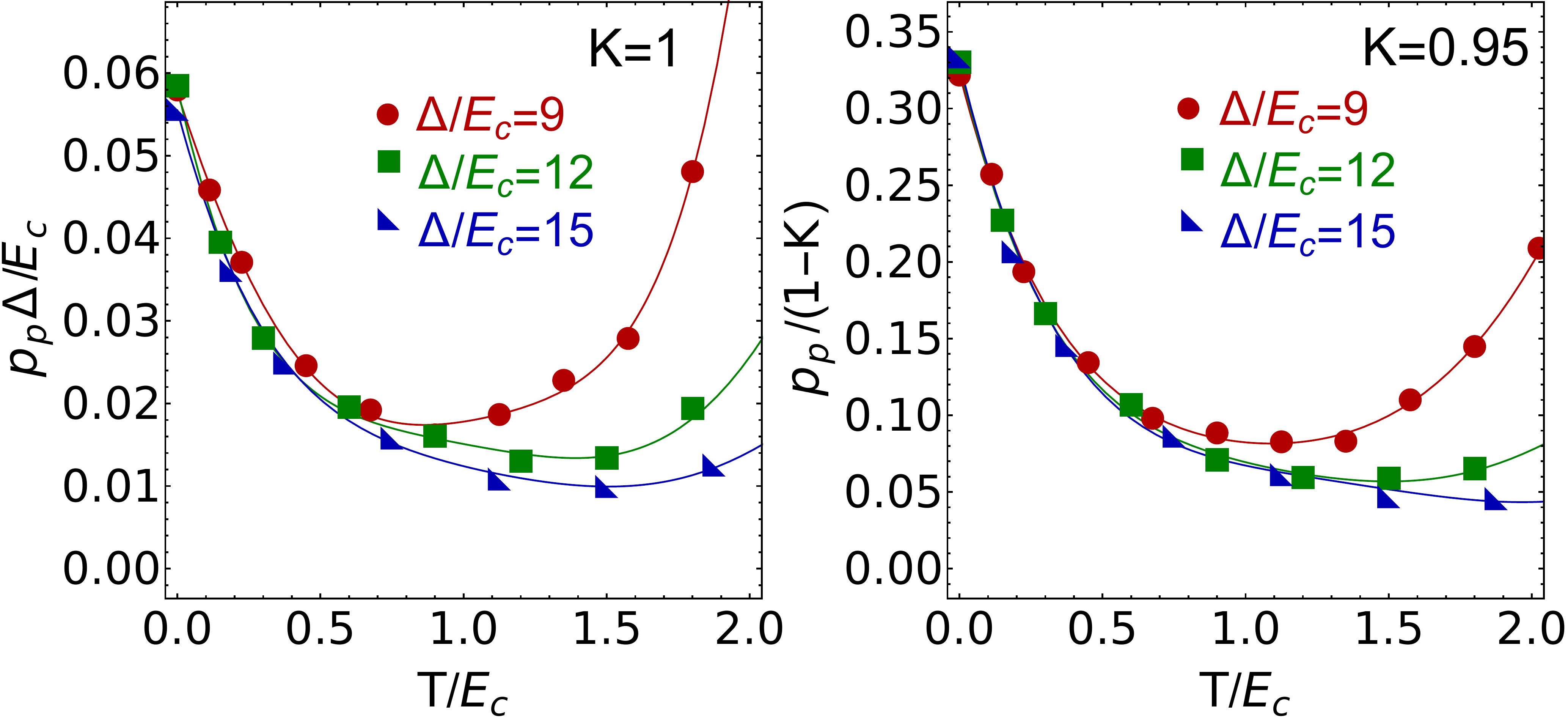}
\caption{Destruction of puddles with increasing temperature.
The fraction $p_p$ (see Fig.~\ref{fig:puddle}) of dopants organized in puddles drops rapidly as function of temperature
at a temperature scale set by the Coulomb interaction $E_c$ between neighbouring dopants.
Numerical results are given for $\gap/E_c$\,=\,9, 12, and 15 (left: perfect compensation, $K$\,=\,1; right: $K$\,=\,0.95).
The scaling plots ($p_p \, \Delta/E_c$ for $K$\,=\,1 and $p_p/(1-K)$ for $1-K \gtrsim 0.3\, E_c/\Delta$, see Fig.~\ref{fig:scaling1},
as function of $T/E_c$) allow us to extrapolate the results to larger values of $\Delta/E_c$.
Scaling demonstrates that the destruction of puddles always occurs at approximately the same value of $T/E_c$.
}
\label{fig:scaling2}
\end{figure}

At $T$\,=\,0, we reproduce the results of Ref.~\cite{Skinner12}, minimizing the energy by a pairwise exchange of charges.
We extend, however, the simulation to finite temperatures using a Monte Carlo approach (see, e.g., Ref.\ \cite{Sarvestani95}
for finite $T$ simulations for other Coulomb systems).
Puddles are formed from occupied donor states or empty acceptor states, see Fig.\ \ref{fig:puddles}.
These correspond to \textit{neutral} dopants, where, e.g., an electron compensates a positively charged donor ion.
To detect puddles in our simulation, we distinguish extended regions of neutral dopants from isolated sites.
Around each neutral dopant, we consider a sphere of radius $r_0$ and count the number $n$ of other neutral dopants of the same type,
see the sketch in Fig.\ \ref{fig:puddle}.
The radius $r_0=1.42\,d_{\rm def}$ is chosen such that on average there are $12$ dopants (the number of nearest neighbours for close-packed spheres)
of the same type within the sphere.
With $p_0(n)$ we denote the fraction of dopants which are neutral and have $n$ neutral neighbours.
This fraction is plotted in Fig.\ \ref{fig:puddle} for perfect compensation $K$\,=\,1 (left panel) as well as for $K$\,=\,0.95 (right).
For $T$\,=\,0, most neutral dopants are organized in puddles, i.e., have many neutral neighbours.
Note, however,  that $p_0(n)$ at $T$\,=\,0 is peaked
at a value of $n$ substantially smaller than 12 which implies that an impurity band formed by donor or acceptor states within a puddle is only partially filled.
This is related to the fact that the energy scale governing puddle formation
(the depth of the potential $E_+(\vec r)-\mu$ in Fig.~\ref{fig:puddles}) is given by $E_c$ and is of the same size as the fluctuations of energy of neighbouring dopants.

Upon increasing $T$, the total density of neutral dopants $(N_A+N_D)\,\sum_{n=0}^\infty p_0(n)$ increases.
For low $T$ this increase is due to the thermal activation of carriers \textit{outside} of the puddles,
as $p_0(n)$ rises only for small number $n$ of neighbours, see Fig.\ \ref{fig:puddle}.
At the same time, the number of neutral dopants with large $n$ decreases, thus also the number of carriers \textit{inside} the puddles decreases.
The reason is that the thermally activated charges screen the Coulomb potential, which leads to a pronounced reduction
of the fluctuations of the Coulomb potential, see Fig.~\ref{fig:puddles}, and therefore to the destruction of puddles
in a highly non-linear process. As we will show, this mechanism of puddle destruction is remarkably robust against
changes of $\Delta/E_c$ and deviations from perfect compensation, $K=1$.
The identification of this mechanism and of the corresponding energy scales is the main result of our theoretical analysis.

To quantify this effect, we count the fraction $p_p=\sum_{n\ge n_0} p_0(n)$ of neutral dopants located well within a puddle.
For the discussion, we choose $n_{0}$\,=\,4, i.e., we count those neutral dopants with at least four neutral neighbours,
as indicated by the vertical dashed lines in Fig.\ \ref{fig:puddle}.
We have also used other values of $n_0$ (not shown) to confirm that the results do not depend qualitatively on this choice.
A similar approach to identify clusters in random systems is, for example, used in Ref.~\cite{clusters15}.

We first consider the limit $T \to 0$. A scaling collapse of all results is obtained when
$p_p \cdot \Delta/E_c$ is plotted as function of $(1-K)\, \Delta/E_c$, see Fig.\ \ref{fig:scaling1}.
For $K$\,=\,1, the density of neutral dopants contributing to puddles is of the order $N_{\rm def}\, E_c/\Delta$ with a small prefactor.
This density does, however, rapidly increase when small deviations from perfect compensation are taken into account.
This occurs roughly when $|N_D-N_A|$, the uncompensated part of the doping, becomes larger than the density of neutral dopants
organized in puddles at $K$\,=\,1.
The linear growth of $p_p$ with $|1-K|$ corresponds to the effect that adding, e.g., a small amount of extra donors to the perfectly compensated system
does \textit{not} give rise to a uniform doping but instead increases the number of neutral donors organized in puddles.
This non-uniform doping originates again in the large-scale inhomogeneities of the Coulomb potential and is important also for the behavior at finite temperature.

Upon increasing $T$, the fraction $p_p$ of dopants in puddles drops sharply, as shown in Fig.~\ref{fig:scaling2}.
For perfect compensation, plots of $p_p(T) \cdot \Delta/E_c$ versus $k_B T/E_c$ are independent of $\Delta/E_c$
for $T, E_c \ll \gap$ (left panel of Fig.~\ref{fig:scaling2}). This allows us to predict that
the Coulomb interaction $E_c$ between neighbouring dopants sets the temperature scale for the destruction of puddles
even in the experimentally relevant regime $\Delta/E_c\sim 100$ (see below).
The right panel of Fig.~\ref{fig:scaling2}, where $p_p/(1-K)$ is investigated for $K$\,=\,0.95,
shows that this physics is not affected by small deviations from perfect compensation.
In the investigated parameter regime ($|1-K|\ll 1$ and $\Delta \gg E_c$), we find that the temperature scale for the destruction
of puddles is independent of doping effects and always given by $E_c$. This has to be contrasted with the strong effects
of doping on the absolute value of $p_p$ which changes by an order of magnitude, see Fig.~\ref{fig:scaling1}.

While the destruction of puddles with increasing temperature has, to our knowledge, not been investigated before,
our  results are fully consistent with known properties \cite{Skinner12,Shklovskii72} of such Coulomb systems.
For $T$\,=\,0 and $K$\,=\,1, it was observed \cite{Skinner12,Shklovskii72} that the density of states of effective single-particle levels
is roughly constant (up to the famous Efros-Shklovskii Coulomb gap at low frequencies) in an energy window set by
$\pm \left(\frac{\gap}{2} + E_c\right)$  and therefore given by $1/\gap$ for $\gap \gg E_c$.
Furthermore, the energy scale of the carriers within a puddle is set by $E_c$.
In combination, this implies that the fraction of charge carriers in puddles is of the order of $E_c/\gap$ in the case of perfect compensation,
as observed in our simulations. This value is, nevertheless, surprisingly large when compared to the much smaller fraction
of charges, $ \sim (E_c/\gap)^3$,  which should be sufficient to compensate charge fluctuations of order $\sqrt{N_{\rm def} R_g^3}$
within the non-linear screening radius $R_g$ discussed in the introduction.

\section{Discussion}
\label{sec:disc}

Both of our main experimental observations, the presence of a sizable optical weight at low $T$ and its rapid drop
on a small temperature scale of the order of 30\,--\,40\,K, agree qualitatively with our numerical simulations
based on a model of shallow donors and acceptors interacting by long-ranged Coulomb interactions.
The remaining task is to compare the parameters of theory and experiment quantitatively.

For the Coulomb energy between neighbouring dopants, theory predicts
\begin{eqnarray}
E_c = \frac{e^2}{4 \pi \epsilon_0 \epsilon} N_{\rm def}^{1/3} \approx k_B \cdot 20 - 40\,{\rm K} \, , \label{ec}
\end{eqnarray}
where we used $\epsilon\approx 200$ (see Fig.\ S4 in \textit{Supplemental Material} \cite{Suppl})
and assumed that the density of shallow donors and acceptors is in the range of
$N_{\rm def}  \sim 10^{19} - 10^{20}\,$cm$^{-3}$ as estimated from the carrier density
in uncompensated samples, see Tab.\ I.\@
This is fully consistent with the experimentally observed temperature scale of 30\,--\,40\,K, which translates to a ratio of
$\Delta/E_c \approx 75-100$.
This agreement does not only corroborate the assertion that puddles exist in BiSbTeSe$_2$
but also confirms the scenario that they arise from strong fluctuations of the Coulomb potential.
Moreover, this agreement indicates that the optical determination of $E_c$ may turn out to be a useful tool to
estimate the defect density $N_{\rm def}$, a quantity which is difficult to assess in a compensated semiconductor.
This interesting result will have to be tested in future experiments.

A quantitative prediction for the effective carrier density $N_{\rm eff}$
is more subtle as this quantity depends for large $\Delta/E_c$ sensitively on the precise amount of compensation which is not known experimentally.
For perfect compensation, $K$\,=\,1, our results for $p_p$ show that $N_p/N_{\rm def}$ should take values of the order of $0.1~E_c/\Delta\sim 10^{-3}$.
The defect density $N_{\rm def}$ can be estimated from the experimental value for $E_c$ and Eq.~(\ref{ec}) which gives $N_{\rm def}$\,=\,$5 - 10 \cdot 10^{19}$\,cm$^{-3}$.
Combining this with the experimental estimate $N_p\,\approx\,2.4\cdot 10^{17}$\,cm$^{-3}$ at 5\,K (see above), we obtain
$N_p/N_{\rm def}\approx 0.002 - 0.005$, a factor 2$-$10 larger than the theoretical estimate for $K$\,=\,1.
Most likely, this just means that compensation is not perfect in our sample. As can be seen from Fig.~\ref{fig:scaling1},
tiny deviations from perfect compensation at a level of $1-K \lesssim 1.5 E_c/\Delta$ can easily explain the observed spectral weight.
With $\Delta/E_c \approx 75-100$, a deviation from perfect compensation of just $1$ or $2\%$ is sufficient to obtain a consistent description of the experiment.

Taking both the extremely simplified nature of the theoretical description and the uncertainties in parameters
like the effective mass into account, the quantitative determination of the parameters should perhaps not be taken too literally.
They are, however, highly plausible, suggesting that at low temperatures we achieve agreement between theory and experiment
at least on a semi-quantitative level.

\section{Conclusion}

Our optical conductivity data of the almost perfectly compensated topological insulator BiSbTeSe$_2$
reveal the existence of puddles at low temperatures as well as their destruction on a temperature scale of 30\,--40\,K.\@
Both the spectral weight and the temperature scale agree semi-quantitatively with our numerical simulations based on a model
of shallow donors and acceptors interacting by long-ranged Coulomb interactions.
We have shown that puddles are suppressed by thermally activated charges which screen the Coulomb potential.
The temperature scale of puddle destruction is set by the Coulomb interaction $E_c$ between neighbouring dopants.
This mechanism works both for near-perfect and perfect compensation.

Puddle formation driven by long-ranged Coulomb interactions
is not only of importance in compensated semiconductors but also for other materials with a vanishing density of electronic states
including Dirac matter in two or three dimensions, like graphene \cite{Martin08} or Weyl semimetals.
For the physics of topological insulators, puddle formation in compensated samples has both positive and negative effects.
While the strong fluctuations of the Coulomb potential imply that it is more difficult to reach high bulk resistivities
despite of perfect compensation, they can also help to localize electrons or holes in puddles in situations
where the compensation of donors and acceptors is not perfect.
The surface states of topological insulators can provide extra screening channels, thus suppressing puddle formation close to the surface or in thin samples.
It  will therefore be interesting to study both experimentally and theoretically,
how puddle formation depends on sample thickness and other parameters
and how it interacts with the charge density of the topological surface states.
Controlling puddle formation may turn out to be a key step for further reduction of bulk transport in topological insulators.

\section{Methods}
\label{sec:exp}

\subsection{Samples}
\label{subsec:samples}

The compound BiSbTeSe$_2$ belongs to the family of $A_2B_3$ tetradymites ($A$\,=\,Bi,Sb; $B$\,=\,Te,Se)
showing rhombohedral structure (space group $R\bar{3}m$) \cite{Nakajima63,Ren11} with three quintuple layers per unit cell
stacked along the [111] direction.
Single crystals of BiSbTeSe$_2$ were grown starting from high-purity elements as described in Ref.~\cite{Ren11}.
The crystals were cut into platelets with typical dimensions of $3\times3$\,mm$^2$ within the (111) plane.
Due to the weak van der Waals bonding between quintuple layers, the samples can be cleaved easily along the (111) plane
using adhesive tape. This yields shiny plane-parallel surfaces.

The defect density can be reduced by chalcogen order \mbox{$B^I$\,-\,$A$\,-\,$B^{II}$\,-\,$A$\,-\,$B^I$} within the
quintuple layers \cite{Ren10,Xiong12,Taskin11,Ren11}.
In Bi$_2$Te$_2$Se with Te\,-\,Bi\,-\,Se\,-\,Bi\,-\,Te order \cite{Ren10,Xiong12},
the $B^{II}$ sites are exclusively occupied by Se ions.
In the solid solutions Bi$_{2-x}$Sb$_x$Te$_{3-y}$Se$_y$ with $y \geq 1$, the composition has been optimized with the aim to achieve
full compensation \cite{Ren11,Pan14}. Chalcogen order is preserved to some extent, as shown by x-ray diffraction \cite{Ren11}.
Among these solid solutions, BiSbTeSe$_2$ was reported to show the highest DC resistivity at 2\,K \cite{Ren11}.
It reaches 3\,$\Omega$cm but varies by about a factor of 3 for samples with the same nominal composition.
This can partially be attributed to a thickness dependence \cite{Taskin11,Xu14,Pan14}
related to a finite conductance contribution of the surface but also reflects different defect concentrations \cite{Ren11}.

\subsection{Optical measurements}
\label{subsec:opticalmeas}

Infrared reflectance and transmittance measurements were performed with unpolarized light in the frequency range of 50\,--\,7500\,cm$^{-1}$
(6\,meV\,--\,0.93\,eV) using a Bruker IFS 66v/S Fourier-transform spectrometer equipped with a continuous-flow He cryostat.
The transmittance $T(\omega)$ was recorded at normal incidence with the electric field parallel to the cleavage plane,
while the reflectivity $R(\omega)$ was measured under near-normal incidence.
Additionally, ellipsometric data were obtained using a rotating analyzer ellipsometer (Woollam VASE) equipped with
a retarder between polarizer and sample. The ellipsometric data were collected at room temperature in the photon energy range
of 0.75\,--\,5.5\,eV (6050\,--\,44360\,cm$^{-1}$) for three different angles of incidence (60$^\circ$, 70$^\circ$, and 80$^\circ$).
Reflectance data and ellipsometric data were measured on a sample with a thickness of $d \! \approx \! 1.1$\,mm.
For the transmittance measurements, we started on a sample with a thickness of $d$\,=\,(183\,$\pm$\,5)\,$\mu$m
(see Fig.\ S1 in \textit{Supplemental Material} \cite{Suppl}).
The value of $d$ was determined mechanically using a micrometer screw. For this rather thick sample,
the accuracy of 5\,$\mu$m corresponds to an error of 2.7\,\%.
Subsequently this sample was cleaved several times using adhesive tape, and the transmittance was measured successively
on the same sample for a series of different thicknesses, $d$\,=\,183, 130, and 102\,$\mu$m. The latter two values
were determined by comparing the Fabry-Perot interference fringes which arise in a transparent
frequency range due to multiple reflections within the sample.
Due to the shiny and plane-parallel surfaces obtained by cleaving, the interference fringes are particularly pronounced,
see Fig.\ S1 in \textit{Supplemental Material} \cite{Suppl}.
For two samples $a$ and $b$ with different thicknesses $d_a$ and $d_b$, the thickness ratio can be determined from
the fringe periods, $\Delta \nu_a/ \Delta \nu_b$\,=\,$d_b/d_a$, with an accuracy of better than 0.5\,\%.
This is important for the comparison of results obtained for different thicknesses,
see Fig.\ S3 in \textit{Supplemental Material} \cite{Suppl}, and thus for the question whether there is a finite contribution of surface states.

In the transparent frequency range, the complex optical conductivity $\tilde{\sigma}(\omega)$\,=\,$\sigma_1(\omega) + i \sigma_2(\omega)$
was determined from $T(\omega)$ and $R(\omega)$ \cite{Grueninger2002}.
In the opaque range, $\tilde{\sigma}(\omega)$ was obtained via a Kramers-Kronig analysis of $R(\omega)$, which at high frequencies was
extrapolated using the ellipsometric results.

\subsection{Monte Carlo simulations}
\label{subsec:simulations}

The model defined in Eq.~(\ref{model}) is simulated at finite $T$ with a standard Monte Carlo algorithm (Metropolis).
Periodic boundary conditions for the Coulomb potential are imposed by using always the shortest distance on the $3-$torus for its computation.
We start the simulations from a configuration where all $N_A$ acceptors and $(1-K)N_D$ donors are occupied, such that the total system is charge neutral.
We only consider configurations which keep charge neutrality by using a pairwise exchange of charge in each Metropolis step.
For $T \to 0$ we average over $100$ disorder realizations.
At high temperatures, averages over only $10$ different realizations turn out to be sufficient.
We simulate up to $N_A+N_D$\,=\,$2N_{\rm def}$\,=\,$2 \times 38^3 \approx 110.000$ dopants.
For the parameters $K$\,=\,1, $T/E_c$\,=\,0, and $\Delta/E_c$\,=\,15 finite-size effects are largest, see \textit{Supplemental Material} \cite{Suppl} for details.
Therefore, we used $2 \times 60^3 = 432.000$ dopants for this particular set of parameters (a single triangle in Figs.~\ref{fig:scaling1} and \ref{fig:scaling2}).

For $T = 0$ the algorithm is identical to the one used in Ref.~\cite{Skinner12}.
It yields only a local and not a global minimum of the (free) energy but such pseudo ground states are known
to describe the properties of real ground states with high accuracy \cite{Skinner12}.
In contrast to simulations with local interactions, the numerical costs for the long-ranged Coulomb interactions increase strongly for increasing temperature and scale with $(N_A+N_D)^2$:
each update of the charge configuration implies that the local energies of all other dopants
have to be recomputed (see Ref.~\cite{Skinner12}).
As puddle formation occurs at length scales of $(  \gap/E_c)^2 /N_{\rm def}^{1/3}$, the number of dopants needed in a simulation
grows with $(  \gap/E_c)^6$ and a worst-case estimate for the computational costs is $(  \gap/E_c)^{12}$ for finite temperatures.

\begin{acknowledgments}
Financial support through the German Excellence Initiative via the key profile area ``quantum matter and materials''
of the University of Cologne is gratefully acknowledged.
The work was also supported by JSPS (KAKENHI 25220708) and MEXT (Innovative Area ``Topological Materials Science'' KAKENHI).
The numerical simulations were performed on the CHEOPS cluster at RRZK Cologne.
\end{acknowledgments}

\clearpage

\begin{centering}
\textbf{Supplemental Material:\\
Revealing puddles of electrons and holes in compensated topological insulators}
\end{centering}

\renewcommand{\theequation}{S\arabic{equation}}
\renewcommand{\thefigure}{S\arabic{figure}}

\setcounter{section}{0}
\setcounter{figure}{0}

\section{Transmittance spectra}

\begin{figure}[b]
\centering
\includegraphics[width=1.0\columnwidth]{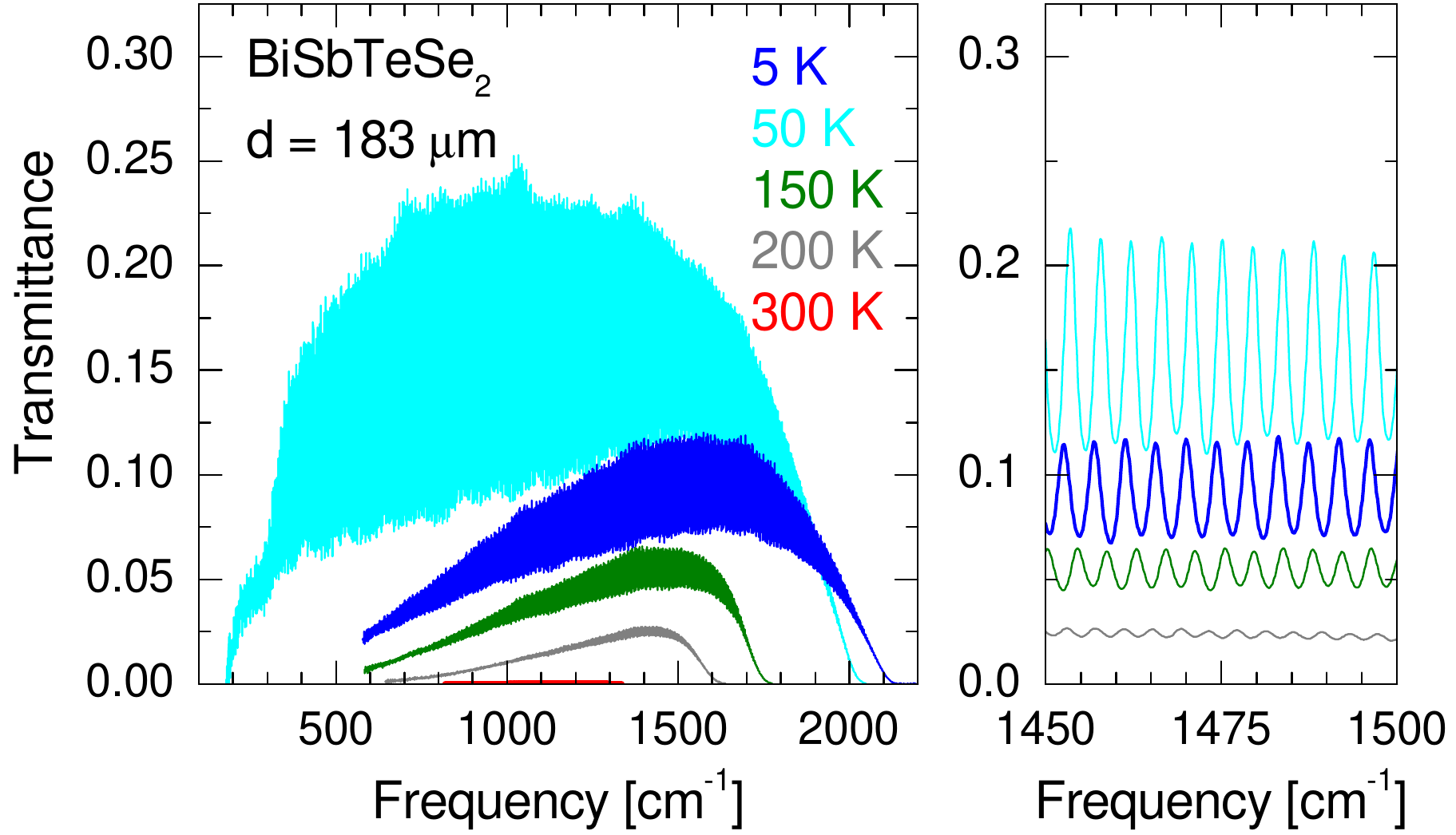}
\caption{Infrared transmittance spectra of BiSbTeSe$_2$.
Left: $T(\omega)$ measured on a sample with a thickness of $d$\,=\,183\,$\mu$m.
Right: same data on an enlarged scale, highlighting the pronounced interference fringes.
}
\label{fig:Trans}
\end{figure}

Figure \ref{fig:Trans} shows the transmittance $T(\omega)$ measured on a sample with a thickness of $d$\,=\,183\,$\mu$m.
At 300\,K, $T(\omega)$ exceeds the noise level only between 830 and 1335\,cm$^{-1}$ and stays below 0.5\%.
However, $T(\omega)$ strongly increases upon cooling down. At 50\,K, the sample is transparent between about 180\,cm$^{-1}$ and 2060\,cm$^{-1}$.
Most remarkably, $T(\omega)$ {\it decreases} upon further cooling below about 50\,K.\@
At 5\,K, $T(\omega)$ is much lower than at 50\,K for frequencies not too close to the gap.
In the highly transparent range, $T(\omega)$ exhibits Fabry-Perot interference fringes (see right panel of Fig.\ \ref{fig:Trans})
caused by multiple reflections between front and back surface.
The fringes are particularly pronounced due the shiny, plane-parallel surfaces of cleaved samples.

\section{Temperature dependence of the gap}

\begin{figure}[b]
\centering
\includegraphics[width=0.9\columnwidth]{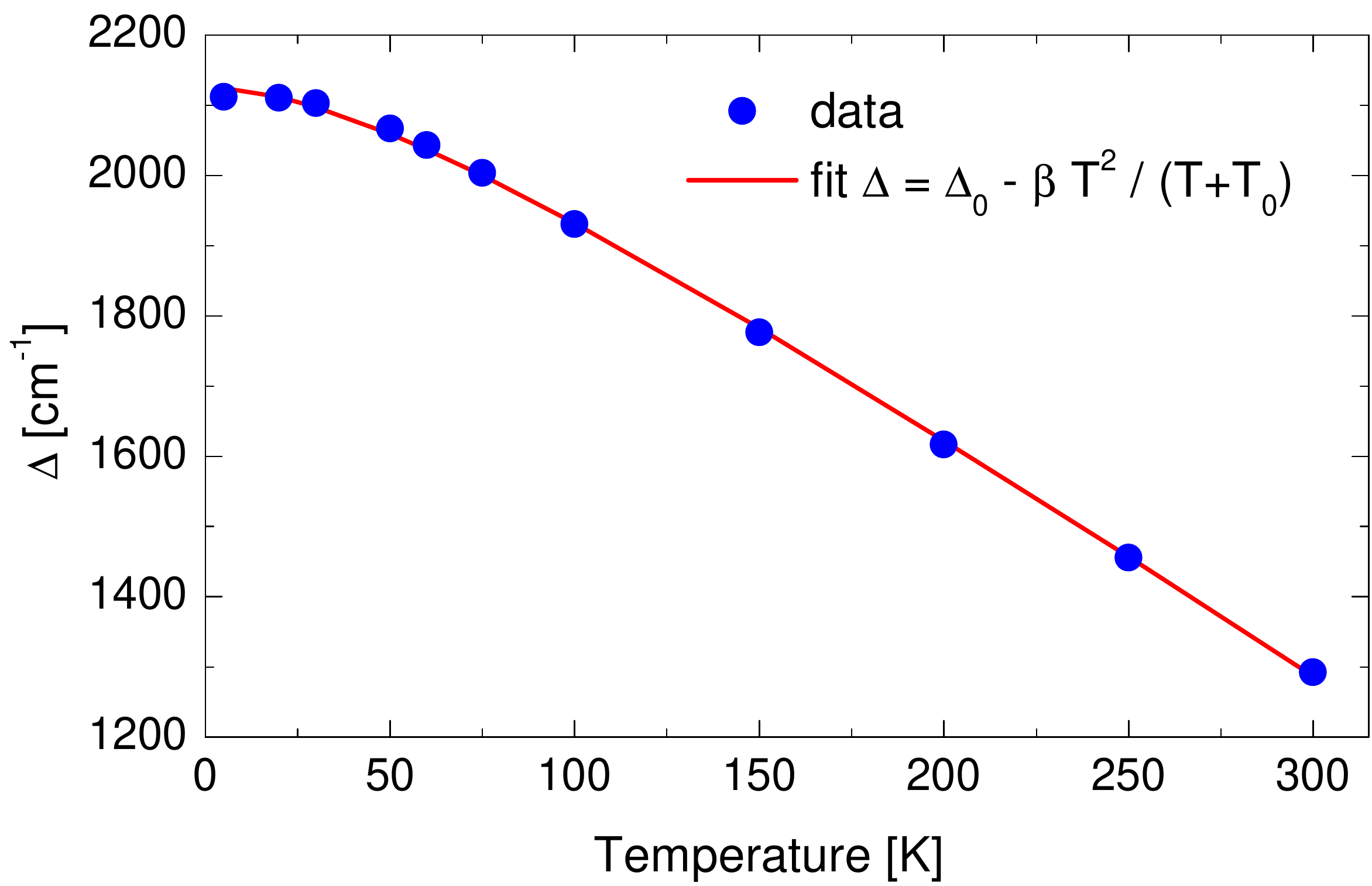}
\caption{Temperature dependence of the gap.
Symbols depict $\Delta(T)$ as determined from the onset of transmittance for a sample thickness of $d$\,=\,102\,$\mu$m.
The red line shows a fit based on the empirical Varshni equation.
}
\label{fig:gap}
\end{figure}

Figure \ref{fig:gap} depicts $\Delta(T)$ as determined from the onset of transmittance for a sample thickness of $d$\,=\,102\,$\mu$m.
The gap shifts by almost 40\% between 5\,K and 300\,K, from about 2112\,cm$^{-1}$ (262\,meV) to 1292\,cm$^{-1}$ (160\,meV).
A fit based on the empirical Varshni equation
$\Delta(T)$\,=\,$\Delta(0) - \beta \cdot T^2/(T+T_0)$ (red line) yields
$\Delta(0)$\,=\,2125\,cm$^{-1}$,
$T_0$\,=\,86\,K, and
$\beta$\,=\,3.6\,cm$^{-1}$/K  or 5.2\,k$_{\rm B}$.
Our result for $\beta$ is framed by values reported for binary compounds,
e.g., 0.77\,cm$^{-1}$/K for Bi$_2$Te$_3$ \cite{sAustin58},
1.6\,cm$^{-1}$/K \cite{sBlack57} and 2.0\,cm$^{-1}$/K \cite{sLaForge10} for Bi$_2$Se$_3$,
and 5.6\,cm$^{-1}$/K for Sb$_2$Se$_3$ \cite{sBlack57}.
Band-structure calculations for binary compounds indicate that $\Delta$ is very sensitive to temperature \cite{sNechaev13,sMichiardi14}.
To the best of our knowledge, band-structure calculations for quaternary compounds have not been reported thus far.

\section{Optical conductivity for different sample thicknesses}

\begin{figure}[t]
\centering
\includegraphics[width=0.95\columnwidth]{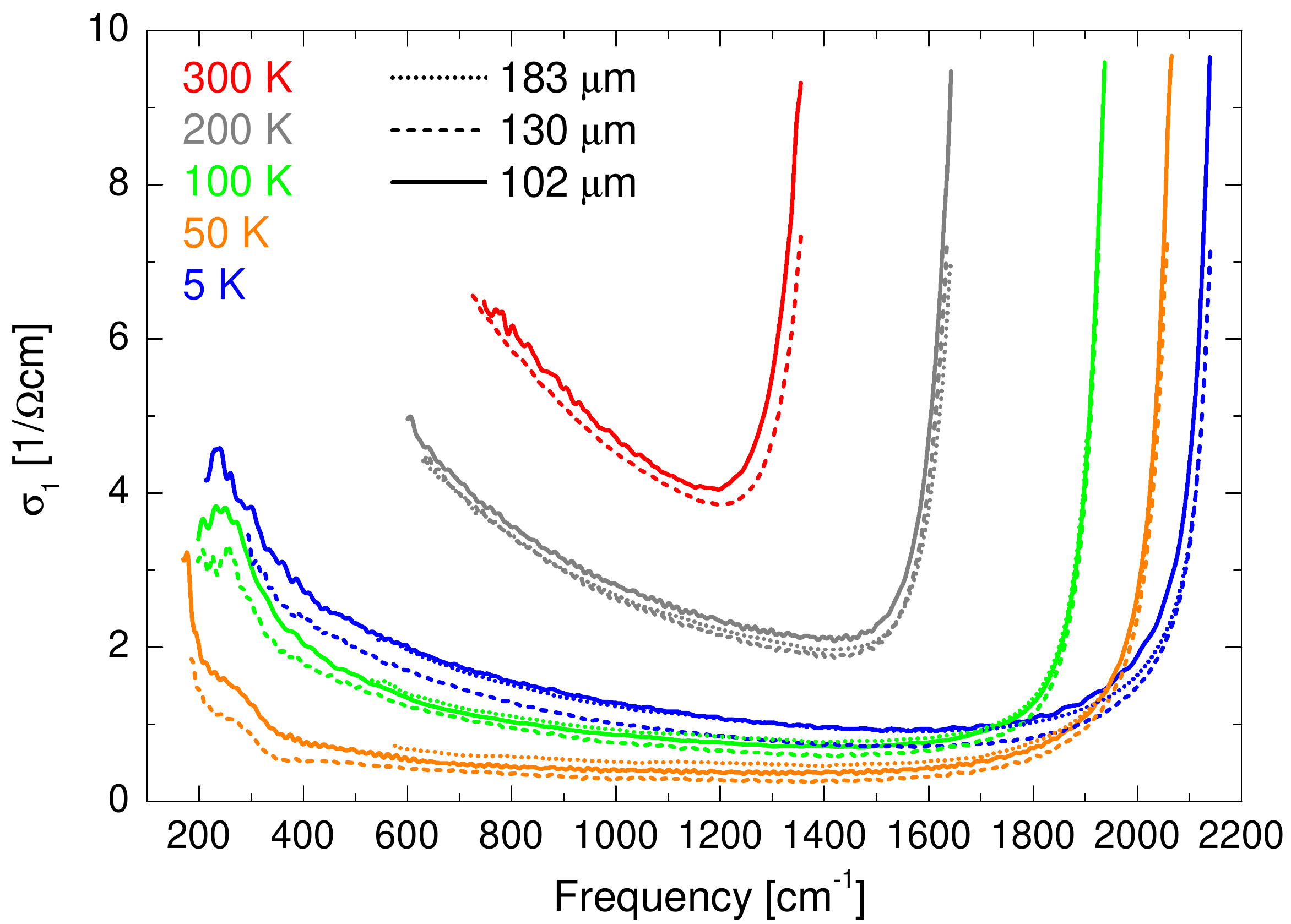}
\caption{Results for the optical conductivity $\sigma_1(\omega)$ for different sample thicknesses $d$ agree very well with each other,
strongly supporting a bulk-only character.
}
\label{fig:compareD}
\end{figure}

The bulk conductivity $\sigma_1(\omega)$ and a possible surface conductance $G(\omega)$ can be disentangled
by comparing results for different thicknesses $d$. If the data analysis is based on the assumption of a bulk-only character,
a finite surface conductance gives rise to an apparent thickness dependence of $\sigma_1(\omega)$,
i.e., the calculated $\sigma_1(\omega)$ increases upon decreasing $d$.
In contrast, our results for $\sigma_1(\omega)$ for different $d$ agree very well with each other, see Fig.\ \ref{fig:compareD}.
At 5\,K, the ratio $\sigma_1^{\rm 102}/\sigma_1^{\rm 183}$ lies within the range 0.99\,--\,1.05 below 1500\,cm$^{-1}$.
This strongly supports a bulk-only character of the investigated excitations.
At 1000\,cm$^{-1}$, the difference between data for $d$\,=\,183\,$\mu$m and 102\,$\mu$m lies in the range $-0.1$\,$\Omega$cm$^{-1}$
to $+0.2$\,$\Omega$cm$^{-1}$ for different temperatures. This can be considered as the experimental uncertainty, which can be attributed
to small experimental errors concerning the absolute value of $T(\omega)$ or the thickness $d$.
In our data, $\sigma_1(\omega)$ typically takes the smallest value for $d$\,=\,130\,$\mu$m.
These values are in particular \textit{smaller} than the results for $d$\,=\,183\,$\mu$m, in contrast to the expectations.
This points towards a small systematic error of the absolute value.

\section{Effective mass}

The effective mass $m^*$ can be estimated by comparing our results with Hall-effect data \cite{sRen11}
which yield a Hall constant of $R_H$\,=\,5\,cm$^3$/C at 300\,K.\@
In a perfectly compensated semiconductor, both electrons and holes contribute to transport.
In this case, $R_H$ gives an upper estimate of the carrier density, $N$\,=\,$1.25\cdot 10^{18}$\,cm$^{-3}$ at 300\,K.\@
However, our analysis (see \textit{Discussion} in main text) suggests that the sample of BiSbTeSe$_2$
shows small deviations of 1\,--\,2\,\% from perfect compensation,
i.e., one carrier type predominates. We thus expect that the result for $N$ derived from $R_H$ is reliable.
From the optical data, we find an effective carrier density $N_{\rm eff}\! \approx \! 6\cdot 10^{18}$\,cm$^{-3}$ at 300\,K,
from which we obtain $m^*/m_e\! \approx \! 0.2$.
In BiSbTeSe$_2$, the cyclotron mass of the bulk bands is unknown since the measured quantum oscillations
arise only from the surface states \cite{sRen11}.
However, the cyclotron mass of the bulk conduction band has been measured in Bi$_2$Se$_3$,
showing $m^*/m_e$ between 0.14 to 0.24 depending on the orientation of the cyclotron orbit \cite{sEto10},
in agreement with our result.

\section{Far-infrared reflectivity}

\begin{figure}[t]
\centering
\includegraphics[width=0.95\columnwidth]{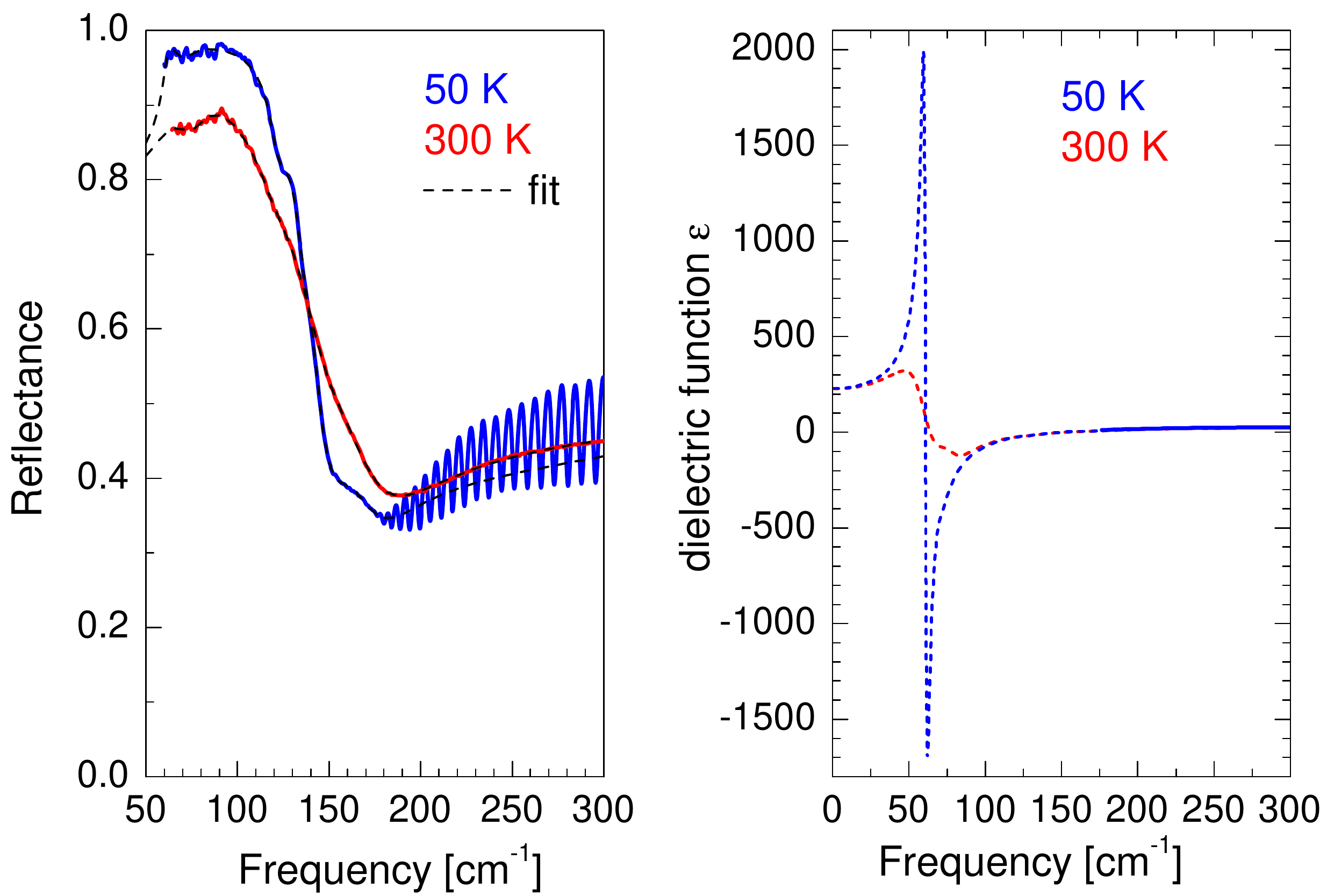}
\caption{Reflectivity and dielectric function $\varepsilon(\omega)$ in the far-infrared range.
Left: Reflectivity for a sample thickness of $d$\,=\,102\,$\mu$m (solid lines) compared to Drude-Lorentz fits (dashed lines).
Right: Dashed lines depict the dielectric function $\varepsilon(\omega)$ determined from the Drude-Lorentz fits
of the reflectivity shown in the left panel. The Reststrahlenband carries a huge oscillator strength.
Solid lines show $\varepsilon(\omega)$ as obtained from the interference fringes in the transparent range.
}
\label{fig:RFIR}
\end{figure}

The reflectivity for a sample thickness of $d$\,=\,102\,$\mu$m is shown in the left panel of Fig.\ \ref{fig:RFIR}.
Infrared-active phonon modes dominate the shape of $R(\omega)$ below about 150\,cm$^{-1}$ (see also
Refs.\ \cite{sLaForge10,sPost15,sReijnders14,sAkrap12,sAleshchenko14,sDiPietro12,sPost13,sDordevic13,sChapler14,sKoehler74ph,sRichter77}).
In the transparent frequency range above the phonons, the measured reflectivity shows interference fringes.
For these measurements with polarization parallel to the cleavage plane, we expect two phonon modes with $E_u$ symmetry \cite{sRichter77}.
These modes were observed at $48$\,cm$^{-1}$ and $98$\,cm$^{-1}$ in Bi$_2$Te$_3$, at $61$\,cm$^{-1}$ and $134$\,cm$^{-1}$
in Bi$_2$Se$_3$ \cite{sRichter77}, and at $62$\,cm$^{-1}$ and $120$\,cm$^{-1}$ in BiSbTeSe$_2$ \cite{sPost15}.
Since the spectral weight of the lower mode is much larger, the reflectivity data typically
show one strong Reststrahlenband above about 50\,cm$^{-1}$ with a weaker feature on its high-frequency side.
In BiSbTeSe$_2$, we find a strong Reststrahlenband with a weak shoulder at about 124\,cm$^{-1}$,
in reasonable agreement with the data reported by Reijnders \textit{et al.} \cite{sReijnders14} and by
Post \textit{et al.} \cite{sPost15}. The strong Reststrahlenband is located close to the lower limit of our frequency range,
which hinders the determination of the precise eigenfrequency.
Compared to Bi$_2$Te$_2$Se, the features are much broader in BiSbTeSe$_2$, which can be attributed
to the Bi/Sb and Te/Se disorder.

\section{Dielectric function $\varepsilon(\omega)$}

For $\omega\to 0$, $\varepsilon(\omega)$ is dominated by the huge oscillator strength of the strong phonon mode,
see right panel of Fig.~\ref{fig:RFIR}. A Drude-Lorentz fit yields $\varepsilon(\omega\to 0)$ of the order of 200.
A more precise determination of this value is difficult because the relevant phonon is close to the lower
frequency limit of our data, see left panel of Fig.~\ref{fig:RFIR}.
Similar values can be derived from results reported in the literature.
In Bi$_2$Te$_2$Se, an analysis of the infrared-active phonon modes yields one strong phonon mode with a contribution of
about 230, six more modes with a total contribution of about 10, and a value of $\varepsilon \! \approx \! 35$\,--\,40
above the phonons \cite{sAkrap12}. In total, this yields $\varepsilon(\omega\to 0) \! \approx \! 275$\,--\,280 in Bi$_2$Te$_2$Se.
In Bi$_2$Se$_3$, a Drude-Lorentz fit yields a phonon contribution of about 170 and a value of
$\varepsilon \! \approx \! 30$ above the phonons, adding up to $\varepsilon(\omega\to 0) \! \approx \! 200$ \cite{sDordevic13}.

\section{Finite-size effects of the numerical data}

\begin{figure}[th]
\centering
\includegraphics[width=0.95\columnwidth]{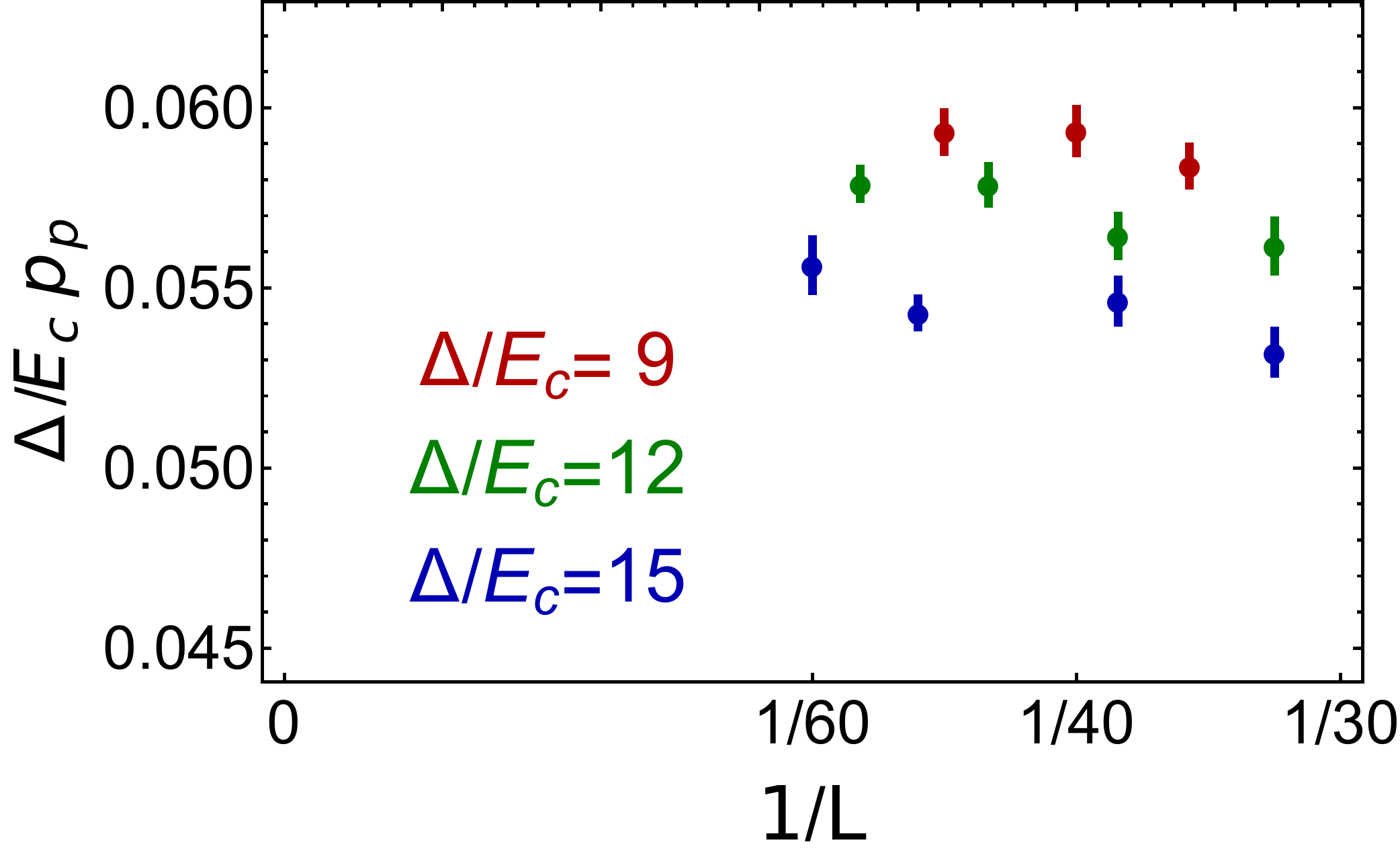}
\caption{To illustrate finite-size effects of our numerical results, $p_p \Delta/E_c$ (see Fig.~7 of the main text) is plotted
as function of the inverse of the linear size $L$ of the simulated box for perfect compensation $K$\,=\,1
and $T$\,=\,0 for $\Delta/E_c$\,=\,9, 12, and 15.
The number of dopants is $2 \times L^3$, reaching more than $400.000$ for $L$\,=\,60. Error bars represent a single standard deviation
of the mean value obtained by averaging over $200-500$ impurity configurations. All finite size effects are well below $5\%$.
\label{finiteSize}
}
\end{figure}

At $T$\,=\,0 and for perfect compensation, $K$\,=1, screening is weakest and one expects the largest finite-size effects in the numerical simulations.
In this regime the typical distance of puddles scales with $(\Delta/E_c)^2$.
In Fig.~\ref{finiteSize} we show $p_p \, \Delta/E_c$, as function of the inverse linear system size,
where $p_p$ denotes the fraction of dopants contributing to electron-hole puddles (see Fig.\ 5 of main text).
While small system-size effects are clearly visible, all finite-size errors are well below $5\%$.
Note that we expect a universal value of  $p_p \,\Delta/E_c$ for $\Delta/E_c \to \infty$.
For the shown values of $\Delta/E_c\ge 9$ this limit is  reached with only small corrections: $p_p \, \Delta/E_c$ slightly grows for smaller values of  $\Delta/E_c$.

\end{document}